\documentstyle[preprint,aps]{revtex}

\def\lsim{\raise0.3ex\hbox{$\;<$\kern-0.75em\raise-1.1ex
\hbox{$\sim\;$}}}
\def\gsim{\raise0.3ex\hbox{$\;>$\kern-0.75em\raise-1.1ex
\hbox{$\sim\;$}}}
\begin{document}
\textheight = 23cm
\topmargin = -1.4cm

\preprint{
\vbox{
\hbox{Fermilab-Pub-02/184-T} \vskip -0.2cm
\hbox{IFT-P.052/2002}\vskip -0.2cm
\hbox{hep-ph/0208163}\vskip -0.2cm
\hbox{August 15, 2002}
}
}
\title{
Parameter Degeneracies in
Neutrino Oscillation Measurement\\ of Leptonic CP and T Violation}

\author{
Hisakazu Minakata$^1$,
Hiroshi Nunokawa$^2$,
Stephen Parke$^3$
}
\address{
$^1$
\sl Department of Physics, Tokyo Metropolitan University \\[-0.3cm]
1-1 Minami-Osawa, Hachioji, Tokyo 192-0397, Japan\\[-0.3cm]
minakata@phys.metro-u.ac.jp
}
\address{
$^2$
\sl Instituto de F\'{\i}sica Te\'orica,
Universidade Estadual Paulista \\[-0.3cm]
Rua Pamplona 145, 01405-900 S\~ao Paulo, SP Brazil\\[-0.3cm]
nunokawa@ift.unesp.br
}
\address{
$^3$
\sl Theoretical Physics Department,
Fermi National Accelerator Laboratory \\[-0.3cm]
P.O.Box 500, Batavia, IL 60510, USA\\[-0.3cm]
parke@fnal.gov
}

\maketitle

\vspace{-0.7cm}
\hfuzz=25pt
\begin{abstract}
The measurement of the mixing angle $\theta_{13}$, 
sign of $\Delta m^2_{13}$ and the CP or T violating phase $\delta$ 
is fraught with ambiguities in neutrino oscillation. 
In this paper we give an analytic treatment of the
paramater degeneracies associated with measuring the $\nu_\mu
\rightarrow \nu_e$ probability and its CP and/or T conjugates.
For CP violation, we give explicit solutions to allow us to 
obtain the regions where there exist
two-fold and four-fold degeneracies. 
We calculate the fractional differences, 
$\left(\Delta \theta /\bar{\theta} \right)$,
between the allowed solutions 
which may be used to compare with the expected sensitivities of 
the experiments. 
For T violation we show that there is always a complete degeneracy 
between solutions with positive and negative $\Delta m^2_{13}$
which arises due to a symmetry
 and cannot be removed by observing 
one neutrino oscillation probability and its T conjugate. 
Thus, there is always a four 
fold parameter degeneracy apart from exceptional points. 
Explicit solutions are also given and the fractional differences 
are computed.
The bi-probability CP/T trajectory diagrams are extensively 
used to illuminate the nature of the degeneracies.

\end{abstract}

\pacs{PACS numbers:14.60.Pq,25.30.Pt}


\section{Introduction}

The discovery of neutrino oscillation in atmospheric neutrino 
observation in Super-Kamiokande \cite {SKatm} and the recent 
accumulating evidences for solar neutrino oscillations 
\cite {solar} naturally suggests neutrino masses and lepton 
flavor mixing. It is also consistent with the result of 
the first man-made beam long-baseline accelerator experiment 
K2K \cite{K2K}.
Given the new realm of lepton flavor mixing whose door is 
just opened, it is natural to seek for a program of 
exploring systematically the whole structure of neutrino 
masses and lepton flavor mixing.

Most probably, the most difficult task in determining the structure 
of lepton mixing matrix, the Maki-Nakagawa-Sakata (MNS) matrix 
\cite{MNS}, is determination of CP violating phase $\delta$ and 
a simultaneous (or preceding) measurement of 
$|U_{e3}| = \sin{\theta_{13}}$. 
We use in this paper the standard notation for the MNS matrix
with $\Delta m^2_{ij} \equiv m^2_j - m^2_i$.
The fact that the most recent analyses of the solar neutrino data 
\cite {solaranalysis} strongly favor the large mixing angle (LMA) 
Mikheev-Smirnov-Wolfenstein (MSW) solution \cite {MSW}
is certainly encouraging for any attempts to measure leptonic 
CP violation.

Since we know that $\theta_{13}$ is small, 
$\sin^2{2 \theta_{13}} \lsim 0.1$, 
due to the constraint imposed by the Chooz reactor experiment 
\cite {CHOOZ} and we do not know how small it is, there is 
two different possibilities. 
Namely, (A) $\theta_{13}$ is determined prior to the experimental 
search for leptonic CP violation, or (B) not. 
The case (A) is desirable experimentally. To determine unknown 
quantities one by one, if possible, is the most sensible way 
to proceed with minimal danger of picking up fake effects. 
But since there is no guarantee that the case (A) {\it is} the case, 
we must prepare for the case (B).
Even in the case (A), experimental determination of 
$\theta_{13}$ always comes with errors, and one must face with the 
similar problem as in the case B within the experimental 
uncertainties.
Moreover, it is known that determination of $\theta_{13}$ in low 
energy conventional super-beam type experiments suffers 
from additional intrinsic uncertainty, the one coming from unknown 
CP violating phase $\delta$. 
See Ref.~\cite {KMN02} for further explanation and a possible way 
of circumventing the problem. 
Therefore, the determination of $\delta$ and $\theta_{13}$ 
are inherently coupled with one another.

Even more seriously, it was noticed by 
Burguet-Castell {\it et al.} \cite{BurguetC} that 
there exist two sets of degenerate solutions 
($\delta_{i}$, $\theta_{13}^{i}$) (i=1, 2) 
even if oscillation probabilities of 
$P(\nu_{\mu} \rightarrow \nu_{e}) \equiv P(\nu)$ 
and its CP conjugate, 
CP$[P(\nu)] \equiv P(\bar{\nu}_{\mu} \rightarrow \bar{\nu}_{e})$ 
is accurately measured.
They presented an approximate but transparent framework of 
analyzing the degeneracy problem, which we follow in this paper.
It was then recognized in Ref.~\cite{MNjhep01} that unknown 
sign of $\Delta m^2_{13}$ leads to a duplication of the ambiguity, 
which entails maximal four-fold degeneracy (see below). 
It was noticed by Barger {\it et al.} \cite {BMW02} that the 
four-fold degeneracy is further multiplied by an ambiguity due to 
approximate invariance of the oscillation probability under 
the transformation  
$\theta_{23} \rightarrow \frac{\pi}{2} - \theta_{23}$. 
A special feature of the degeneracy problem at the oscillation maximum 
was noted and analyzed to some detail \cite{KMN02,BMW02}. 
Recently, the first discussion of the problem of parameter 
degeneracy in T violation measurement is given in 
Ref.~\cite {MNP02}.

Meanwhile, there were some technological progresses in 
analyzing the interplay between the genuine CP phase and
the matter effects in measuring leptonic CP or T violation 
in neutrino oscillation, the issue much-discussed but 
still unsettled \cite {cp-matter,t-matter}. 
The authors of Refs.~\cite{MNjhep01} and 
\cite{MNP02} introduced,  respectively, 
the ``CP and T trajectory diagrams in bi-probability space''
for pictorial representation of 
CP-violating and CP-conserving phase effects as well as 
the matter effect in neutrino oscillations.
They showed that when these two types of trajectory diagrams 
are combined it gives a unified graphical representation of 
neutrino oscillations in matter \cite{MNP02}. 
We demonstrate in this paper that they provide a powerful 
tool for understanding and analyzing the problem of parameter 
degeneracy, as partly exhibited in 
Refs.~\cite{MNjhep01,taup2001.mina,BMW02}.

It is the purpose of this paper to give a completely general treatment 
of the problem of parameter degeneracy in neutrino oscillations 
associated with CP and T violation measurements. 
We elucidate the nature of the degeneracy, and 
determine the region where it occurs, namely, 
the regions in the $P$-CP[$P$] (and $P$-T[$P$]) 
bi-probability space in which the same-$\Delta m^2_{13}$-sign 
and/or the mixed-$\Delta m^2_{13}$-sign degeneracies take place. 
While partial treatment of the parameter degeneracy 
has been attempted for CP measurement before 
\cite{KMN02,BurguetC,MNjhep01,BMW02} such general treatment 
is still lacking. 
We believe that it is worthwhile to have such an overview of the 
parameter degeneracy issue to uncover ways of resolving this
problem. 
See \cite{BMW02b,HLW02,DMM02,BurguetC2} for recent discussions.

We present the first systematic discussion of parameter degeneracy 
in T measurement following our previous paper in which we set up the 
problem \cite{MNP02}. 
We uncover a new feature of the degeneracy in T measurement. 
Namely, we show that for a given T trajectory diagram there always 
exists an another T diagram which completely degenerates with the 
original one and has opposite sign of $\Delta m^2_{13}$. 
It means that for any given values of $P(\nu)$ and T[$P(\nu)$]  
there is two degenerate solutions of ($\delta$, $\theta_{13}$) 
with differing sign of $\Delta m^2_{13}$. 
It should be noticed that this is true no matter how large 
the matter effect, quite contrary to the case of CP measurement. 
Therefore, determination of the signs of $\Delta m^2_{13}$ is 
impossible in a single T measurement experiment 
unless one of the following conditions is met; 
(a) one of the solutions is excluded, for example, by the CHOOZ 
constraint, or 
(b) some additional information, such as energy distribution of 
the appearance electrons, is added.

We emphasize that a complete understanding 
of the structure of the parameter degeneracy should be helpful for 
one who want to pursue solution of the ambiguity problem in an 
experimentally realistic setting. 
We, however, do not attempt to discuss the 
$\theta_{23} \leftrightarrow \frac{\pi}{2} - \theta_{23}$
degeneracy.
We also do not try to solve the problem of parameter degeneracy 
exactly though it is in principle possible by using an exact but 
reasonably compact expression of the oscillation probability 
obtained by Kimura, Takamura, and Yokomakura \cite{KTY02}. 
Instead, we restrict ourselves into the treatment with the 
approximation introduced by Burguet-Castell {\it et al.} 
\cite{BurguetC} 
in which the approximate formula for the oscillation 
probability derived by Cervera {\it et al.} \cite{golden} 
was employed. 
Though not exact, it gives us much more transparent overview of 
the problem of parameter degeneracy.

\section{Problem of parameter ambiguity in CP and T violation 
measurement}

We define the ``CP (T) parameter ambiguity'' as the problem of having 
multiple solutions of ($\delta, \theta_{13}$)
and the sign of $\Delta m^2_{23}$,   
for a given set of measured values of 
oscillation probabilities of 
$P(\nu_{\mu} \rightarrow \nu_{e})$ and its CP (T) conjugate, 
CP$[P(\nu)] \equiv P(\bar{\nu}_{\mu} \rightarrow \bar{\nu}_{e})$ 
(T$[P(\nu)] \equiv P(\nu_{e} \rightarrow \nu_{\mu})$).
We concentrate in this paper on this channel because precise 
measurement is much harder in other channels, e.g., in 
$\nu_{e} \rightarrow \nu_{\tau}$.
Our use of $\nu_{\mu} \rightarrow \nu_{e}$ 
and its CP-conjugate 
is due to our primary concern on conventional super-beam type 
experiments \cite{lowecp}.
The reader should keep this difference in mind 
if they try to compare our equations with those in 
Refs.~\cite{BurguetC,MNP02} 
in which they use $\nu_{e} \rightarrow \nu_{\mu}$ and its 
CP-conjugate, a natural choice for neutrino factories 
\cite{golden,nufact}.
It should also be noted that the neutrino factories and the superbeam
experiments are studying processes which are T-conjugates.

In this section we utilize the CP and the T trajectory 
diagrams introduced in \cite {MNjhep01} and \cite {MNP02}, 
respectively, to explain what is the problem of parameter ambiguity 
and to achieve qualitative understanding of the solutions 
without using equations. But before entering into the business 
we want to justify, at least partly, our setting, i.e., 
prior determination of all the remaining parameters besides 
$\delta$ and $\theta_{13}$.

\subsection{Problem of parameter degeneracy; set up of the problem}

We assume that at the time that an experiment for measuring 
($\delta, \theta_{13}$) is carried out all the remaining 
parameters in the MNS matrix, 
$\theta_{23}$, $|\Delta m^2_{23}|$, 
$\theta_{12}$, and $\Delta m^2_{12}$, are determined accurately. 
The discussion on how the experimental uncertainties 
affect the problem of parameter degeneracy is important, 
but is beyond the scope of this paper. 
It should be more or less true, because $\theta_{23}$ 
and $\Delta m^2_{23}$ will be determined quite 
accurately by the future long-baseline experiments 
\cite {JHF,MINOS,OPERA} up to the 
$\theta_{23} \leftrightarrow \frac{\pi}{2} -\theta_{23}$ ambiguity. 
Most notably, the expected sensitivities in the JHF-SK 
experiment in its phase I is \cite{JHF}
\begin{eqnarray}
&\delta&(\sin^2{2 \theta_{23}}) \simeq 10^{-2}
\nonumber \\
&\delta&(|\Delta m^2_{23}|) \simeq 4 \times 10^{-4} \mbox{eV}^2 
\label{JHFsensitivity}
\end{eqnarray} 
at around $|\Delta m^2_{23}| = 3 \times 10^{-3}$ eV$^2$.
On the other hand, the best place for accurate determination of 
$\theta_{12}$ and $\Delta m^2_{12}$, 
assuming the LMA MSW solar neutrino solution, is most probably the 
KamLAND experiment; the expected sensitivities are 
\cite{KamLANDerror}
\begin{eqnarray}
&\delta&(\tan^2{\theta_{12}}) \simeq 10 \%
\nonumber \\
&\delta&(\Delta m^2_{12}) \simeq 10 \%
\label{KamLDsensitivity}
\end{eqnarray} 
at around the LMA best fit parameters.
Therefore, we feel that our setting, prior determination 
of all the mixing angles and $\Delta m^2$'s besides $\delta$, 
$\theta_{13}$ and the sign of $\Delta m^2_{23}$ , 
is reasonable one at least in the first 
approximation.

\subsection{Pictorial representation of parameter ambiguities}

In this subsection we use CP and the T trajectory diagrams 
\cite {MNjhep01,MNP02} to explain intuitively what is the problem 
of parameter degeneracy, and to achieve qualitative understanding 
of the solutions without using equations.
In Fig.~1 we display four CP trajectories in the 
$P$-CP[$P$] bi-probability space which all have intersection 
at $P(\nu)$ = 1.9\% and CP[$P(\nu)$] = 2.6\%.
The four CP trajectories are drawn with four different values 
of $\theta_{13}$, 
$\sin^2{2 \theta_{13}} = 0.055$, 0.050, 0.586, 0.472, and 
the former (latter) two trajectories correspond to positive 
(negative) $\Delta m^2_{13}$. 
The analytic expressions of the four degenerate solutions will be 
derived in Sec.~IV. 
The neutrino energy $E$ and the baseline length $L$ are taken as 
$E=250$ MeV and $L=130$ km, respectively\footnote{
In all the figures of this paper we do not average over
neutrino energy distributions as was performed in Refs.~\cite{MNjhep01,MNP02}
but use a fixed neutrino energy specified for each figure.}. 
The setting anticipates 
an application to the CERN-Frejus project \cite {CERNFR}, 
where the regions with parameter ambiguities are widest. 
The values of all the remaining mixing parameters are given in the 
caption of Fig.~1.

\begin{figure}[t]
\vspace*{15cm}
\includegraphics{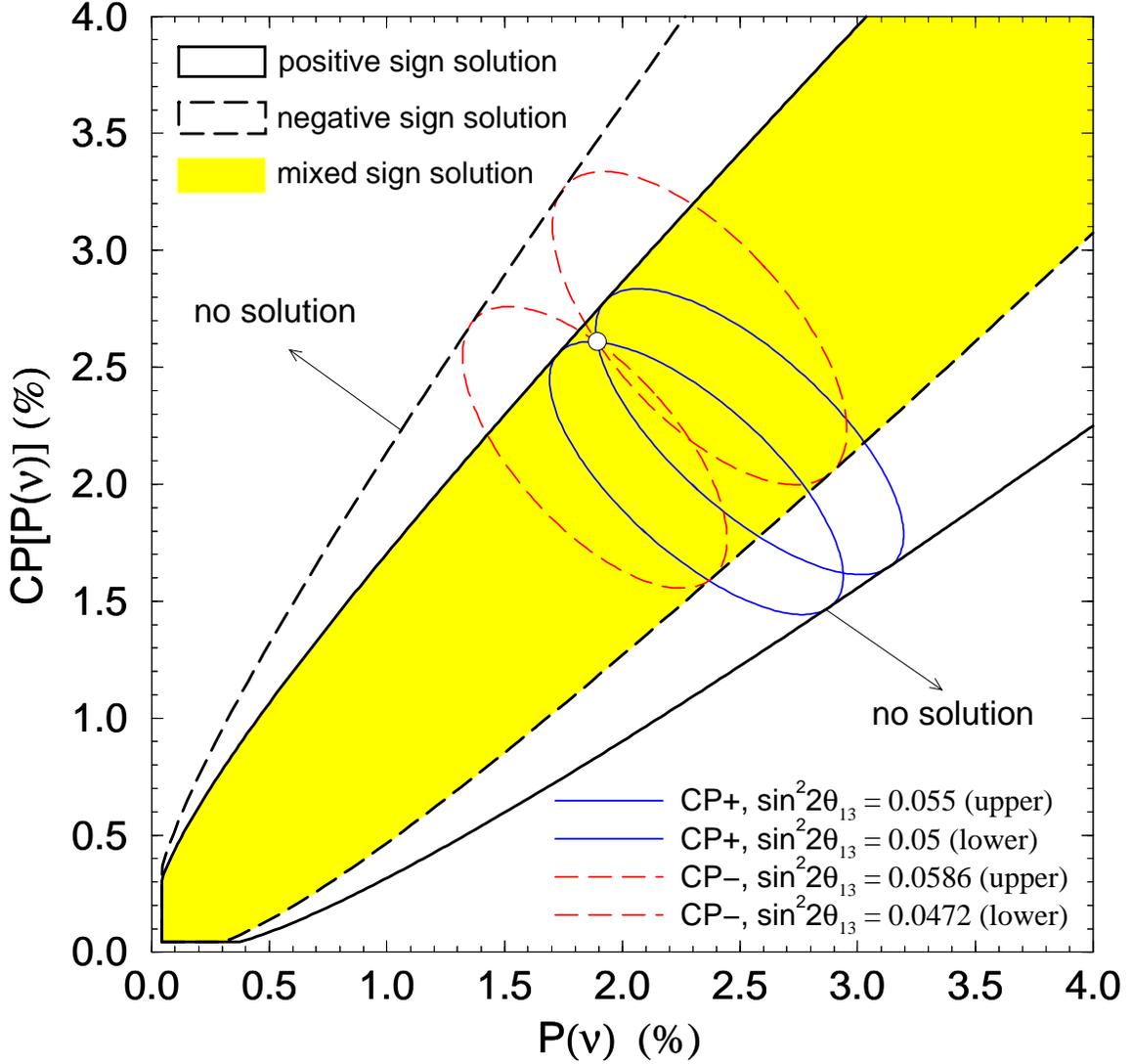}
\vspace{1.0cm}
\caption[]{
An example of the degenerate solutions for the CERN-Frejius project
in the $P(\nu) \equiv P(\nu_\mu \rightarrow \nu_e)$ verses 
$CP[P(\nu)] \equiv P(\bar{\nu}_\mu \rightarrow \bar{\nu}_e)$
plane.  
Between the solid (dashed) lines is the allowed region for
positive (negative) $\Delta m^2_{13}$ and the shaded region 
is where solution for both signs are allowed. 
The solid (dashed) ellipses are for positive (negative) $\Delta m^2_{13}$
and they all meet at a single point.
This is the CP parameter degeneracy problem.
We have used a fixed neutrino energy of 250 MeV and 
a baseline of 130 km.  
The mixing parameters are fixed to be 
$|\Delta m^2_{13}| = 3 \times 10^{-3} eV^2$,
$\sin^2 2\theta_{23}=1.0$,
$\Delta m^2_{12} = +5 \times 10^{-5} eV^2$,
$\sin^2 2\theta_{12}=0.8$ and
$ Y_e \rho  = 1.5$ g cm$^{-3}$.
}
\label{CPegfig}
\end{figure}

Figure 1 demonstrates that there is a four-hold degeneracy 
in the determination of ($\delta$, $\theta_{13}$) 
for a given set of $P(\nu)$ and CP[$P(\nu)$]. 
The region between the solid lines and the region between the dashed
lines in Fig.~1 
are the regions where two-fold degeneracies exist in the 
solutions of ($\delta$, $\theta_{13}$) 
for positive and negative $\Delta m^2_{13}$ sectors, respectively. 
(See Eq.~(\ref{condition}) in Sec.~IV.) 
It is intuitively understandable that the region 
where degenerate solutions exists is the region swept 
over by the CP trajectories when the parameter $\theta_{13}$ 
is varied, while keeping other mixing parameters and the 
experimental conditions fixed.  
The shaded region is the region where the full four-hold degeneracy exists.

Now we turn to the T measurement. 
In Fig.~2 we display four T trajectories in the 
$P$-T[$P$] bi-probability space which all have intersection 
at $P(\nu)$ = 1.7\% and T[$P(\nu)$] = 2.5\%.
The four T trajectories are drawn with four different values 
of $\theta_{13}$, 
$\sin^2{2 \theta_{13}} = 0.05$, 0.0427, 0.575, 0.490, and 
the former (latter) two trajectories correspond to positive 
(negative) $\Delta m^2_{13}$. 
Matter effects split the positive and negative $\Delta m^2_{13}$
trajectories, see \cite{MNP02}, thus different values of 
$\theta$ are required for mixed sign trajectories to overlap.
The remaining mixing parameters and 
the experimental setting are the same as in Fig.~1. 

\begin{figure}[t]
\vspace*{16cm}
\includegraphics{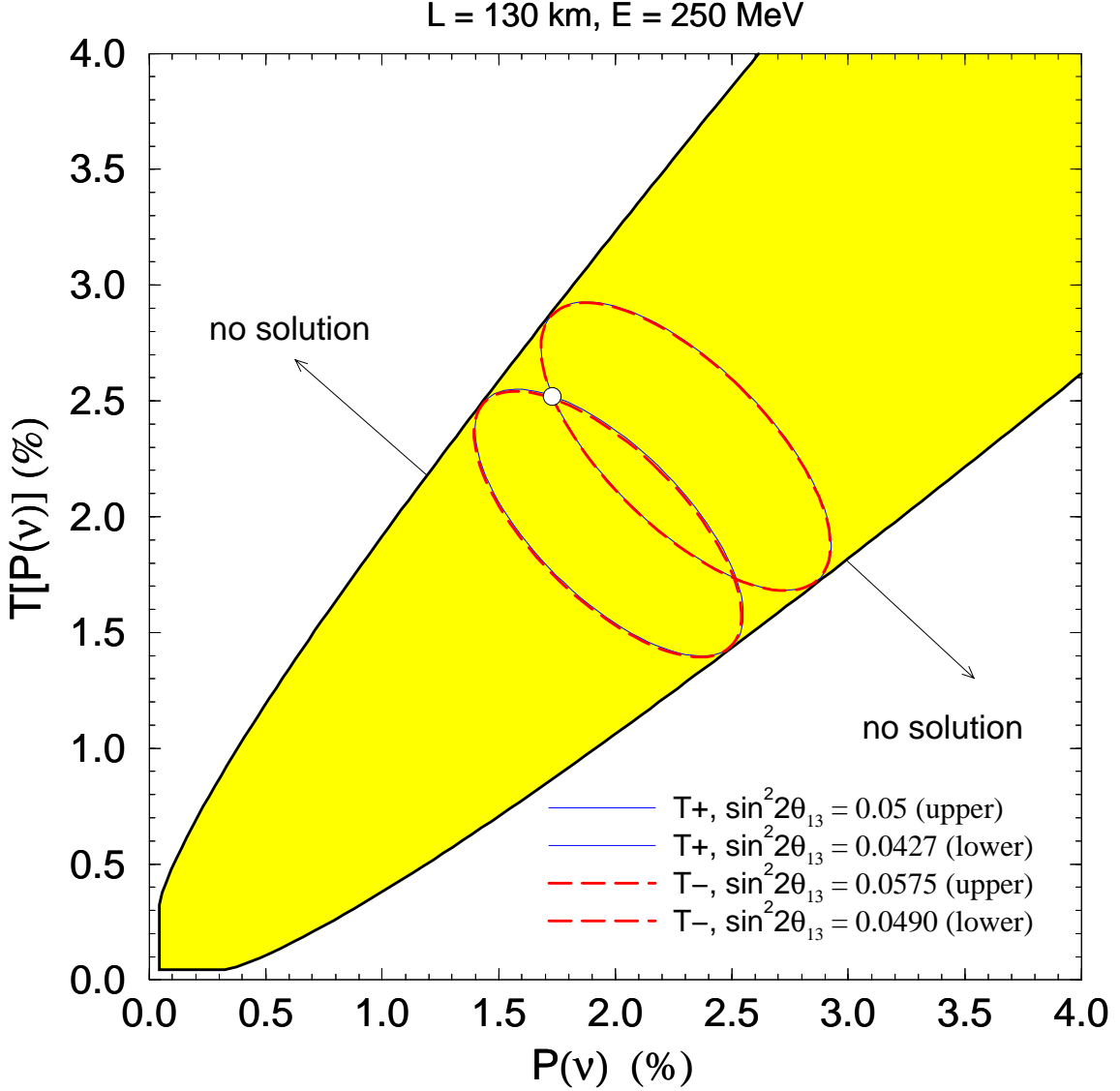}
\vspace{1.0cm}
\caption[]{
An example of the degenerate solutions using the energy and path length of
the CERN-Frejius project
in the $P(\nu) \equiv P(\nu_\mu \rightarrow \nu_e)$ verses 
$T[P(\nu)] \equiv P(\nu_e \rightarrow \nu_\mu)$
plane.  
For this figure there is complete overlap in the region (shaded) 
that allows solutions for either sign of $\Delta m^2_{13}$.
The solid (dashed) ellipses are for positive (negative) $\Delta m^2_{13}$
and they all meet at the ``measured point'', (P, P$^T$) = (1.7, 2.5)\%.
This is the T-parameter degeneracy problem.
Notice that for the each ellipse with positive  $\Delta m^2_{13}$
there is a completely degenerate ellipse with negative $\Delta m^2_{13}$.
This feature will be explained in Sec. IIIc.
Parameters are the same as in Fig.~\ref{CPegfig}.}
\label{Tegfig}
\end{figure}

A clear and interesting difference from the CP diagram manifests 
itself already at this level, reflecting the highly symmetric 
nature of the T conjugate probabilities as will be made explicit in 
eq.~(\ref{Tequation}).
Namely, the two different-$\Delta m^2_{13}$-sign diagrams 
(the first and the third, and the second and fourth) 
completely overlap with each other. 
In the next section we will make it clear that the complete 
degeneracy originates from a symmetry.
Therefore, discrimination of the sign of $\Delta m^2_{13}$ 
is impossible in a single T-violation measurement experiment
unless one of the solutions is excluded by an other experiment.
We will demonstrate in Sec.~III that 
the degeneracy is not accidental one specific to this particular 
case, but its existence is generic. 
There is always a four-hold degeneracy in T measurement.

\section{Parameter degeneracy in T-violation measurements}

We start by presenting an analytic treatment of the problem of parameter 
degeneracy in T-violation measurements primarily because it is
simpler and instructive.  
To do this 
we generalize the formalism developed by 
Burguet-Castell {\it et al.} \cite{BurguetC} by treating 
the cases of positive and negative $\Delta m^2_{13}$ 
simultaneously.
It provides basis of our treatment of the degeneracy between 
the same as well as across the alternating 
$\Delta m^2_{13}$-sign probabilities. 
It will become clear from the following discussions 
that the treatment of the mixed-sign degenerate solutions,
for both CP and T measurements, 
can be done as a straightforward generalization of the 
same-sign degeneracy case by simply taking account of 
duplication due to the alternating sign of $\Delta m^2_{13}$ 
\cite{MNjhep01}.

There are four basic equations 
satisfied by the T-conjugate probabilities in the case of 
T measurement for small $\sin \theta_{13}$: 
\begin{eqnarray}
P(\nu)_{+} &=& X_{+} \theta^2 + 
Y_{+} \theta \cos {\left( \delta + \frac{\Delta_{13}}{2} \right)} + 
P_{\odot}
\nonumber \\
T[P(\nu)]_{+} &\equiv& P^T(\nu)_{+} = 
X_{+} \theta^2 + 
Y_{+} \theta \cos {\left( \delta - \frac{\Delta_{13}}{2} \right)} +  
P_{\odot} 
\nonumber \\
P(\nu)_{-} &=& X_{-} \theta^2 + 
Y_{-} \theta \cos {\left( \delta - \frac{\Delta_{13}}{2} \right)} + 
P_{\odot}
\nonumber \\
T[P(\nu)]_{-} &\equiv& P^T(\nu)_{-} = 
X_{-} \theta^2 + 
Y_{-} \theta \cos {\left( \delta + \frac{\Delta_{13}}{2} \right)} + 
P_{\odot}
\label{Tequation}
\end{eqnarray}
where $X_{\pm}$ and $Y_{\pm}$ are given in Appendix, 
$P_{\odot}$ indicates the term which is related with 
solar neutrino oscillations, and 
$\Delta_{13} \equiv \frac{|\Delta m^2_{13}| L}{2E}$. 
Note that $\pm$ here refers to the sign of $\Delta m^2_{13}$ 
and $\theta$ is an abbreviation of $\theta_{13} \simeq \sin \theta_{13}$.
In the next subsections we discuss 
the possible solutions for $\theta$ and $\delta$ for a given 
measurement of both P and $P^T$ for both
positive and negative sign of $\Delta m^2_{13}$.

\subsection{The same-sign degeneracy; T measurement}

The treatment in this subsection applies
for two overlapping T trajectories with the same sign of 
$\Delta m^2_{13}$. The degeneracy associated with 
alternating-sign trajectories will be explored in the next subsection.

There are two sets of approximate solutions
of (\ref{Tequation}), 
$\theta_{i}$ and $\delta_{i}$ , where  
(i = 1,2) and (i = 3,4) denotes the solutions in the 
positive and negative $\Delta m^2_{13}$ sectors, respectively. 
They are 
\begin{eqnarray}
\theta_{i} &=& \sqrt{\frac{P-P_{\odot}}{X_{\pm}}} - 
\frac{Y_{\pm}}{2X_{\pm}} 
\cos {\left( \delta_{i} \pm \frac{\Delta_{13}}{2} \right)} 
\nonumber \\
\theta_{i} &=& \sqrt{\frac{P^T - P_{\odot}}{X_{\pm}}} - 
\frac{Y_{\pm}}{2X_{\pm}} 
\cos {\left( \delta_{i} \mp \frac{\Delta_{13}}{2} \right)} 
\end{eqnarray}
where $\pm$ correspond to solutions in positive and negative 
$\Delta m^2_{13}$ sectors
\footnote{The above solutions are exact solutions to the system of Eq. 
(\ref{Tequation}) if we were to add terms  
$Y^2_\pm \cos^2 \left(\delta \pm \frac{\Delta_{13}}{2} \right)/(4X_\pm)$
to the equations, (\ref{Tequation})
. In what follows we have systematically ignored terms of 
${\cal{O}}(Y^2_\pm/X_\pm)$.}. 
We then obtain, e.g., for the 
positive $\Delta m^2_{13}$ sector 
\begin{eqnarray}
\theta_{2} - \theta_{1} &=& - 
\frac{Y_{+}}{2X_{+}} 
\left[
\left(\cos{\delta_{2}} - \cos{\delta_{1}} \right)
\cos{\left(\frac{\Delta_{13}}{2}\right)} - 
\left(\sin{\delta_{2}} - \sin{\delta_{1}} \right)
\sin{\left(\frac{\Delta_{13}}{2}\right)} 
\right] \nonumber \\
\theta_{2} - \theta_{1} &=& - 
\frac{Y_{+}}{2X_{+}} 
\left[
\left(\cos{\delta_{2}} - \cos{\delta_{1}} \right)
\cos{\left(\frac{\Delta_{13}}{2}\right)} + 
\left(\sin{\delta_{2}} - \sin{\delta_{1}} \right)
\sin{\left(\frac{\Delta_{13}}{2}\right)} 
\right] 
\label{Tsolsame1}
\end{eqnarray}
which entails the degeneracy that if $(\theta_1,\delta_1)$ is a solution
so is 
\begin{equation}
\theta_{2} =  \theta_{1} + 
\frac{Y_{+}}{X_{+}} 
\cos{\delta_{1}}
\cos{\left(\frac{\Delta_{13}}{2}\right)}
\hskip 0.5cm \mbox{and} \hskip 0.5cm 
\delta_{2} = \pi - \delta_{1},
\end{equation}
in addition to the trivial solution. A similar degeneracy holds
also for the negative $\Delta m^2_{13}$ sector, that is, if
$(\theta_3,\delta_3)$ is a solution so is
\begin{equation}
\theta_{4} =  \theta_{3} + 
\frac{Y_{-}}{X_{-}} 
\cos{\delta_{3}}
\cos{\left(\frac{\Delta_{13}}{2}\right)}
\hskip 0.5cm \mbox{and} \hskip 0.5cm 
\delta_{4} = \pi - \delta_{3}.
\end{equation}
Both of these same sign  $\Delta m^2_{13}$ degeneracies
are in matter though they look 
like the vacuum degeneracies as discussed in \cite{MNjhep01}.
Notice that if the experimental setup is chosen such that
$\cos{\left(\frac{\Delta_{13}}{2}\right)} =0$ \cite{KMN02} or nature has chosen
$\cos{\delta}=0$ then the same sign degeneracies are removed. 

\subsection{The mixed-sign degeneracy; T measurement}

Let us now examine the problem of parameter degeneracy 
which involves positive and negative $\Delta m^2_{13}$. 
The basic equations (\ref{Tequation}) can be approximately solved for 
mixed sign situation as:
\begin{eqnarray}
\theta_{1} &=& \sqrt{\frac{P-P_{\odot}}{X_{+}}} - 
\frac{Y_{+}}{2X_{+}} 
\cos {\left( \delta_{1} + \frac{\Delta_{13}}{2} \right)}
\nonumber \\
\theta_{1} &=& \sqrt{\frac{P^T-P_{\odot}}{X_{+}}} - 
\frac{Y_{+}}{2X_{+}} 
\cos {\left( \delta_{1} - \frac{\Delta_{13}}{2} \right)}
\label{Tmixed1} \\
\theta_{3} &=& \sqrt{\frac{P-P_{\odot}}{X_{-}}} - 
\frac{Y_{-}}{2X_{-}} 
\cos {\left( \delta_{3} - \frac{\Delta_{13}}{2} \right)}
\nonumber \\
\theta_{3} &=& \sqrt{\frac{P^T-P_{\odot}}{X_{-}}} - 
\frac{Y_{-}}{2X_{-}} 
\cos {\left( \delta_{3} + \frac{\Delta_{13}}{2} \right)}.
\label{Tmixed2} 
\end{eqnarray}
We will now exactly determine $\theta$ and $\delta$ 
using the above set of approximate solutions, (\ref{Tmixed1}) and 
(\ref{Tmixed2}), as our starting point. 
First,
\begin{eqnarray}
\sin{\delta_{1}} \sin {\frac{\Delta_{13}}{2}} &=&  
-\frac{\sqrt{X_{+}}}{Y_{+}}
\left(
\sqrt{P - P_{\odot}} - \sqrt{P^T-P_{\odot}}
\right)\\
\sin{\delta_{3}} \sin {\frac{\Delta_{13}}{2}} &=& 
\frac{\sqrt{X_{-}}}{Y_{-}}
\left(
\sqrt{P - P_{\odot}} - \sqrt{P^T-P_{\odot}}
\right) 
\end{eqnarray}
and $\cos{\delta_{i}}$ is given by
$\cos{\delta_{i}} = \pm \sqrt{1-\sin^2{\delta_{i}}}$.
Using these $\cos \delta_i$ the values of $\theta$ are  given by
\begin{eqnarray}
\theta_{1}  &=& \frac{\left(\sqrt{P^{~}-P_{\odot}}+\sqrt{P^T-P_{\odot}}\right)}
{2\sqrt{X_{+}}}
-\frac{Y_{+}}{2X_{+}} \cos\delta_1 \cos{\frac{\Delta_{13}}{2}}
\\
\theta_{3}  &=& \frac{\left(\sqrt{P - P_{\odot}}+\sqrt{P^T - P_{\odot}}\right)}
{2\sqrt{X_{-}}}
-\frac{Y_{-}}{2X_{-}} \cos\delta_3 \cos{\frac{\Delta_{13}}{2}}.
\end{eqnarray}
To relate these alternating sign solutions we use the identity,
see Appendix, 
\begin{equation}
\frac{\sqrt{X_{+}}}{Y_{+}} = - \frac{\sqrt{X_{-}}}{Y_{-}}
\label{keyequation}
\end{equation}
derivable under the 
Cervera {\it et al.} approximation\footnote{
Unless Eq.~(\ref{keyequation}) holds we get into trouble  
because then Eqs.~(\ref{Tmixed1}) or (\ref{Tmixed2}) does 
not allow the (same-sign) solution $\theta_{1}=\theta_{2}$ and 
$\delta_{1}=\delta_{2}$, which must exist as shown in 
Ref.~\cite{MNP02}. Therefore, use of the formula of oscillation 
probability obtained by Cervera {\it et al.} who summed up all order 
matter effect is essential.
}.
Then, it follows that 
\begin{eqnarray}
\sin{\delta_{1}} &=& \sin{\delta_{3}} 
\nonumber \\
(\cos{\delta_{1}} + \cos{\delta_{3}} )
\cos {\frac{\Delta_{13}}{2}} 
&=& -2 
\frac{\sqrt{X_{+}}}{Y_{+}}
\left(
\sqrt{X_{+}} \theta_{1} -  
\sqrt{X_{-}} \theta_{3}
\right)
\label{Tsolmixed1}
\end{eqnarray}
One can choose without loss of generality 
$\delta_{3} = \pi - \delta_{1}$ as a solution of 
(\ref{Tsolmixed1}). 
Then, for a given $P$ and $P^T$ measurement, apart from $(\theta_1,\delta_1)$
there are three other solutions\footnote{As a convention, we have chosen 
$\cos \delta_1 \leq 0$ so that for $\Delta_{13} \leq \pi$, 
$\theta_1 \geq \theta_2$ and $\theta_3 \geq \theta_4$.}
given by
\begin{eqnarray}
\theta_{2} &=& \theta_{1} +
\frac{Y_{+}}{X_{+}} 
\cos{\delta_{1}} \cos {\left(\frac{\Delta_{13}}{2}\right)}
\hskip 0.5cm \mbox{and} \hskip 0.5cm 
\delta_{2} = 
\pi - \delta_{1}
\nonumber \\
\theta_{3} &=& 
\sqrt{\frac{X_{+}}{X_{-}}} 
\theta_{1}
\hskip 3.5cm \mbox{and} \hskip 0.5cm 
\delta_{3} = 
\pi - \delta_{1}
\label{Tsol}
\\
\theta_{4} &=& \theta_{3} -
\frac{Y_{-}}{X_{-}} 
\cos{\delta_{1}} \cos {\left(\frac{\Delta_{13}}{2}\right)}
\hskip 0.5cm \mbox{and} \hskip 0.5cm 
\delta_{4} = \delta_{1}
\nonumber 
\end{eqnarray}
Therefore, there is no ambiguity in determination 
of $\delta$ in T violation measurement 
apart from the one $\delta \rightarrow \pi - \delta$
independent of the sign of $\Delta m^2_{13}$.
Fortunately, this degeneracy does not obscure existence 
or non-existence of leptonic T (or CP) violation. 
This feature arises because of highly constrained nature of  
system (\ref{Tequation}) of T-conjugate probabilities.

The physically allowed region of the T diagram is determined by 
the constraint that
$\sin^2 \delta_i \leq 1$  which in terms of $P$ and $P^T$ is
\begin{equation}
\left(
\sqrt{P - P_{\odot}} - \sqrt{P^T-P_{\odot}}
\right)^2 
\quad \leq \quad \frac{Y^2_{+}}{X_{+}}
\sin^2 {\left(\frac{\Delta_{13}}{2}\right)} 
= \frac{Y^2_{-}}{X_{-}} \sin^2 {\left(\frac{\Delta_{13}}{2}\right)} 
\end{equation}
and is the same region for both signs of
$\Delta m^2_{13}$ because of the identity Eq.(\ref{keyequation}).
In Fig. \ref{Tegfig}, this region is the shaded region using
the CERN-Frejius parameters.
At the boundary of the allowed physical region $\cos \delta = 0$
and the same sign degeneracy vanishes, however, the opposite sign
degeneracy is non-zero. 

We define the fractional differences, 
${\Delta \theta}/{\bar{\theta}}$, by 
\begin{equation}
\left(\frac{\Delta \theta}{\bar{\theta}}\right)_{ij} 
\equiv 
{ \theta_i-\theta_j \over (\theta_i + \theta_j)/2 },
\label{FracDiff}
\end{equation}
to quantify how different the two degenerate solutions are. 
In fact, one can obtain simple expressions for the various fractional 
differences;
\begin{eqnarray}
\left(\frac{\Delta \theta}{\bar{\theta}}\right)_{12} 
& = & \left(\frac{\Delta \theta}{\bar{\theta}}\right)_{34} 
={  - 2 Y_{+} \cos{\delta_{1}} \cos {\left(\frac{\Delta_{13}}{2}\right)}
\over \sqrt{X_{+}}
\left(
\sqrt{P - P_{\odot}} + \sqrt{P^T-P_{\odot}}
\right) },
\label{D12}
 \\
\left(\frac{\Delta \theta}{\bar{\theta}}\right)_{31} 
&=&\left(\frac{\Delta \theta}{\bar{\theta}}\right)_{42} 
= 2{(\sqrt{X_+} - \sqrt{X_-}) \over (\sqrt{X_+} + \sqrt{X_-}) },
 \\
\left(\frac{\Delta \theta}{\bar{\theta}}\right)_{14} 
& \approx &
{  -2Y_{+} \cos{\delta_{1}} \cos {\left(\frac{\Delta_{13}}{2}\right)}
\over \sqrt{X_{+}}
\left(
\sqrt{P - P_{\odot}} + \sqrt{P^T-P_{\odot}} 
\right) }
- 2\left( { \sqrt{X_+} -\sqrt{X_-} \over \sqrt{X_+} + \sqrt{X_-}  } \right),
\\
\left(\frac{\Delta \theta}{\bar{\theta}}\right)_{32} 
& \approx &
{  -2Y_{+} \cos{\delta_{1}} \cos {\left(\frac{\Delta_{13}}{2}\right)}
\over \sqrt{X_{+}}
\left(
\sqrt{P - P_{\odot}} + \sqrt{P^T-P_{\odot}} 
\right) }
+ 2\left( { \sqrt{X_+} -\sqrt{X_-} \over \sqrt{X_+} + \sqrt{X_-}  } \right).
\label{D32}
\end{eqnarray}
The same sign fractional difference, (1,2) and (3,4), 
decreases with increasing $P$ and $P^T$
and thus $\theta$, whereas the first mixed sign fractional difference, 
(3,1) and (4,2),
is independent of the size of $P$ and $P^T$ and thus $\theta$.
The second and third mixed sign fractional differences, (1,4) and (2,3), are 
similar to the
same sign fractional difference but off set by an energy dependent constant.
The relationship between the fractional difference in the measured quantity 
$\sin^2 2\theta$ and $\theta$
is simply
\begin{equation}
{\sin^2 2\theta_i - \sin^2 2\theta_j 
\over
(\sin^2 2\theta_i + \sin^2 2\theta_j)/2 }
\approx
2 {\theta_i - \theta_j 
\over
(\theta_i + \theta_j)/2} 
= 2\left(\frac{\Delta \theta}{\bar{\theta}}\right)_{ij} .
\end{equation}
for $1 \gg  \theta_i, \theta_j
\gg |\theta_i - \theta_j|$.

In Fig.~3 thru 5 we have plotted the differences in the allowed
$\theta$ solutions divided by half the sum for the 
CERN-Frejus, JHF-SK and FNAL-NuMI\cite{NuMI} possible experiments using
$\nu_\mu \rightarrow \nu_e$ and its T-conjugate $\nu_e \rightarrow \nu_\mu$.
The regions where this fractional difference is small are regions
where the parameter degeneracy inherent in such measurements is only
important once the experimental resolution on $\theta$ for a 
fixed solution is of the same size or smaller.
Notice that near the boundaries on the allowed region
the fractional differences are small for the same sign solutions.
For the mixed sign, either the (1,4) or (3,2) fractional difference plots 
have a line for which the fractional difference is zero.  
This line can be understood as follows;
for a given small value of $\theta$ the positive and negative 
$\Delta m^2_{13}$ ellipses overlap and intersect at two points.
As $\theta$ varies these intersection points give us this line
with zero mixed sign fractional difference.

\begin{figure}[t]
\vspace*{9.5cm}
\includegraphics{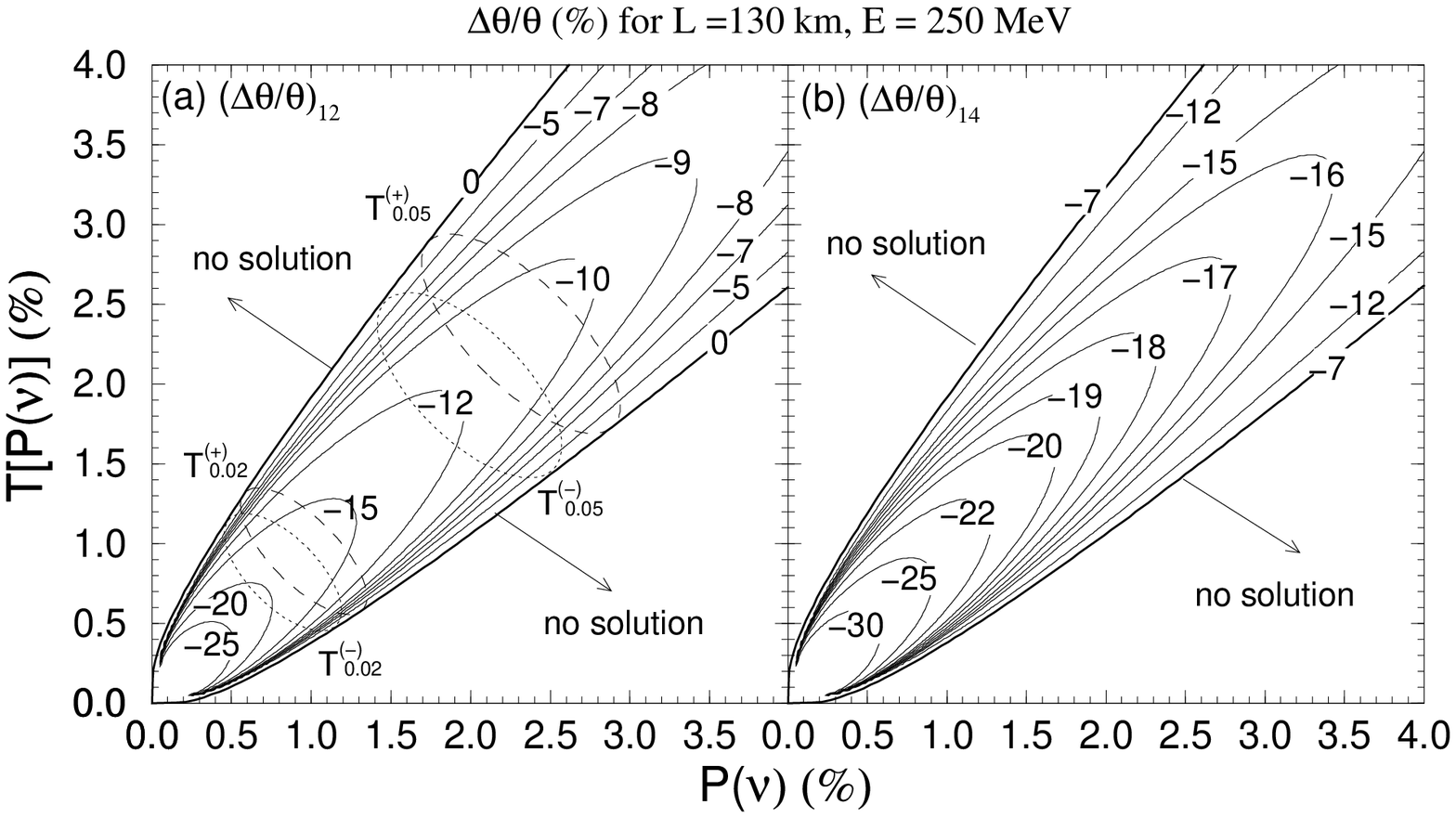}
\vspace{1.0cm}
\caption[]{
The iso-fractional differences, as a \%, for the allowed solutions
for the mixing angle $\theta_{13}$ in the 
$P(\nu) \equiv P(\nu_\mu \rightarrow \nu_e)$ verses 
$T[P(\nu)] \equiv P(\nu_e \rightarrow \nu_\mu)$ plane
for the CERN-Frejius project.
The fractional differences for solutions (3,4) is identical to that for (1,2)
and the fractional difference for (3,2) equals the fractional difference
for (1,4) plus or minus a constant, see eq.~(\ref{D12})-(\ref{D32}). 
In this case the fractional difference (3,2) has a zero contour.
The parameters are the same as in Fig.~\ref{CPegfig}.
The ellipses are labelled $T^{(\pm)}_{\sin^2 2 \theta_{13}}$ 
to show the relevant
size of $\sin^2 2 \theta_{13}$ for this figure.
}
\label{T-CERN}
\end{figure}

\begin{figure}[t]
\vspace*{9cm}
\includegraphics{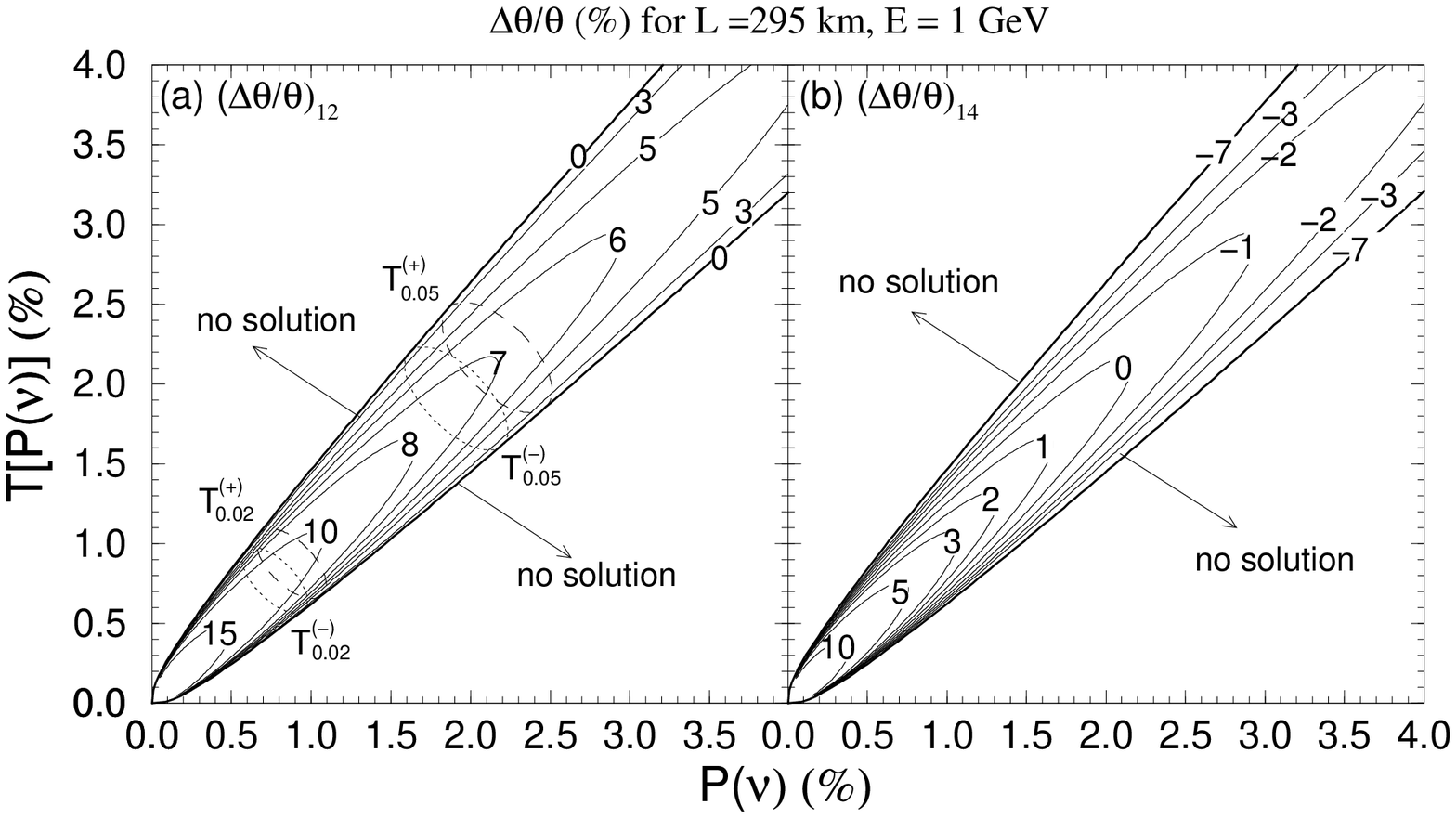}
\vspace{1.0cm}
\caption[]{
The iso-fractional differences, as a \%, for the allowed solutions
for the mixing angle $\theta_{13}$ in the 
$P(\nu) \equiv P(\nu_\mu \rightarrow \nu_e)$ verses 
$T[P(\nu)] \equiv P(\nu_e \rightarrow \nu_\mu)$ plane
for the JHK-SK project.
The fractional differences for solutions (3,4) is identical to that for (1,2)
and the fractional difference for (3,2) equals the fractional difference
for (1,4) plus or minus a constant, see eq.~(\ref{D12})-(\ref{D32}). 
The zero contour appearing in the (1,4) fractional difference is
explained in the text.
The parameters are the same as in Fig.~\ref{CPegfig}.
The ellipses are labelled as in Fig.\ref{T-CERN}.}
\label{T-JHF}
\end{figure}

\begin{figure}[t]
\vspace*{9cm}
\includegraphics{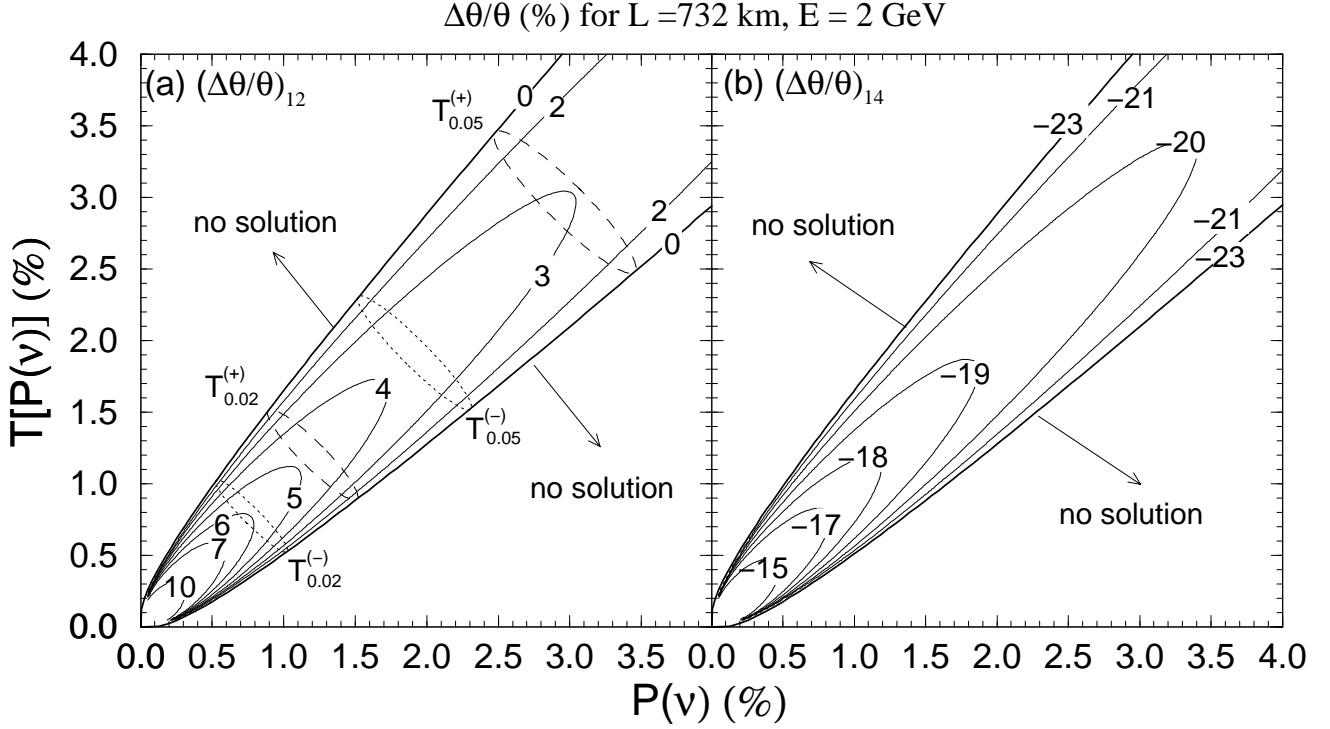}
\vspace{1.0cm}
\caption[]{
The iso-fractional differences, as a \%, for the allowed solutions
for the mixing angle $\theta_{13}$ in the 
$P(\nu) \equiv P(\nu_\mu \rightarrow \nu_e)$ verses 
$T[P(\nu)] \equiv P(\nu_e \rightarrow \nu_\mu)$ plane
for the FNAL-NUMI project.
The fractional differences for solutions (3,4) is identical to that for (1,2)
and the fractional difference for (3,2) equals the fractional difference
for (1,4) plus or minus a constant, see eq.~(\ref{D12})-(\ref{D32}). 
At very small probability the (1,4) fractional difference has a zero contour.
The parameters are the same as in Fig.~\ref{CPegfig}.
The ellipses are labelled as in Fig.\ref{T-CERN}.}
\label{T-NUMI}
\end{figure}

\subsection{Symmetry between the two alternating-$\Delta m^2_{13}$-sign 
T diagrams}

The observant reader will notice that there is in a general one-to-one 
correspondence between the solutions with positive  $\Delta m^2_{13}$,
labelled 1 and 2, 
and those solutions with negative  $\Delta m^2_{13}$, 
labelled 3 and 4, in Eq.~(\ref{Tsol}) by using the identity,
 Eq.~(\ref{keyequation}).
In fact this correspondence applies not only to the
solutions but to the complete T diagram. 
For a given T trajectory with positive $\Delta m^2_{13}$ 
there always exists a T trajectory with negative $\Delta m^2_{13}$ 
with a different value of $\theta$, 
which nevertheless completely overlaps the positive trajectory,
see Fig.~\ref{Tegfig}.
This surprising phenomenon occurs because there exists a 
symmetry in T probability system defined in Eq.~ (\ref{Tequation}).

One can observe from eq.~(\ref{Tequation}) that a positive 
$\Delta m^2_{13}$ trajectory which is defined by the first 
two equations of (\ref{Tequation}) can be used to generate 
a negative $\Delta m^2_{13}$ trajectory which completely 
overlaps with the original one by the transformation 
\begin{eqnarray}
\delta &\rightarrow& \pi - \delta 
\nonumber \\
\theta &\rightarrow& 
\sqrt{\frac{X_{-}}{X_{+}}} \theta 
\label{highsym}
\end{eqnarray}
This means that for a measure set of $P$ and $P^T$ there is always 
two set of solutions with different sign of $\Delta m^2_{13}$. 
There is no way to resolve this ambiguity because the two 
T trajectories are completely degenerate. 
It should be noticed that this situation occurs no matter how 
large a matter effect at much longer baseline. 
In such a case, two T trajectories with the same $\theta_{13}$ 
but opposite sign of $\Delta m^2_{13}$ are far apart, and one 
would have expected that there is no ambiguity 
in determination of the sign of $\Delta m^2_{13}$. 
Hence, there is a ``no-go theorem'' for the determination of 
the sign of $\Delta m^2_{13}$ by a single T violation measurement. 
The possible cases in which the ``theorem'' is circumvented are, 
as mentioned in Introduction,  
(a) one of the solutions is excluded, for example, by the CHOOZ 
constraint, or 
(b) some additional information, such as energy distribution of 
the appearance electrons
or another T-violation measurement with different parameters, 
is added.

\section{Parameter degeneracy in CP-violation measurements}

We now turn to the analytic treatment of the parameter 
degeneracy in CP-violation measurements. 
We proceed in an analogous way to the analytic treatment of
T-violation given in sec. III.

We start with the four basic CP equations for small $\sin \theta_{13}$: 
\begin{eqnarray}
P(\nu)_{+} &=& X_{+}\theta^2 + 
Y_{+} \theta \cos {\left( \delta + \frac{\Delta_{13}}{2} \right)} + 
P_{\odot} \nonumber \\
\mbox{CP}[P(\nu)]_{+} &\equiv& \bar{P}(\nu)_{+} = 
\bar{X}_{+} \theta^2 + 
\bar{Y}_{+} \theta \cos {\left( \delta - \frac{\Delta_{13}}{2} \right)} + 
P_{\odot} \nonumber \\
P(\nu)_{-} &=& X_{-} \theta^2 + 
Y_{-} \theta \cos {\left( \delta - \frac{\Delta_{13}}{2} \right)} + 
P_{\odot} \nonumber \\
\mbox{CP}[P(\nu)]_{-} &\equiv& \bar{P}(\nu)_{-} = 
\bar{X}_{-} \theta^2 + 
\bar{Y}_{-} \theta \cos {\left( \delta + \frac{\Delta_{13}}{2} \right)} + 
P_{\odot}
\label{CPequation}
\end{eqnarray}
where $X_{\pm}$ and $Y_{\pm}$ are given in Appendix. 
As before, $P_{\odot}$ indicates the term which is related with 
solar neutrino oscillations, 
$\Delta_{13} \equiv \frac{|\Delta m^2_{13}| L}{2E}$,
the $\pm$ here refers to the sign of $\Delta m^2_{13}$ 
and $\theta$ is an abbreviation of $\theta_{13} \simeq s_{13}$.

Note that there exist relations among coefficients;
\begin{eqnarray}
X_{\pm} &\equiv& X(\pm \Delta m^2_{13}, a) \\
\bar{X}_{\pm} &=& X(\pm \Delta m^2_{13}, -a) 
\label{symmetry1}
\end{eqnarray}
In leading order in $\frac{\Delta m^2_{12}}{\Delta m^2_{13}}$, 
there exist further relations, 
\begin{equation}
X_{+} = \bar{X}_{-} \hskip 0.5cm \mbox{and} \hskip 0.5cm 
X_{-} = \bar{X}_{+} 
\label{symmetry2}
\end{equation}
which follows from the CP-CP relation \cite{MNP02} (see Appendix). 
Finally, it follows under the approximation of Cervera {\it et al.} 
\cite{golden} that 
\begin{equation}
Y_{+} = - \bar{Y}_{-}, \hskip 0.5cm Y_{-} = - \bar{Y}_{+}.
\label{symmetry3}
\end{equation}

We fully utilize the symmetry relationships 
(\ref{symmetry2}) and (\ref{symmetry3}) in the unified 
treatment of the same-sign and the mixed-sign 
degeneracies. 
The basic equations (\ref{CPequation}) can be solved for 
generic mixed sign situation as: 
\begin{eqnarray}
\theta_{1} &=& \sqrt{\frac{P-P_{\odot}}{X_{+}}} - 
\frac{Y_{+}}{2X_{+}} 
\cos {\left( \delta_{1} + \frac{\Delta_{13}}{2} \right)}\\
\theta_{1} &=& \sqrt{\frac{\bar{P}-P_{\odot}}{X_{-}}} + 
\frac{Y_{-}}{2X_{-}} 
\cos {\left( \delta_{1} - \frac{\Delta_{13}}{2} \right)}\\
\theta_{3} &=& \sqrt{\frac{P-P_{\odot}}{X_{-}}} - 
\frac{Y_{-}}{2X_{-}} 
\cos {\left( \delta_{3} - \frac{\Delta_{13}}{2} \right)}\\
\theta_{3} &=& \sqrt{\frac{\bar{P}-P_{\odot}}{X_{+}}} + 
\frac{Y_{+}}{2X_{+}} 
\cos {\left( \delta_{3} + \frac{\Delta_{13}}{2} \right)}
\end{eqnarray}

The solution of these equations are:
\begin{eqnarray}
&\sin{\delta_{1,2}}& = \frac{1}{D}
\left[
-C^{(+)}
\sin {\frac{\Delta_{13}}{2}} \Delta P_{+}  
\pm
C^{(-)}
\cos {\frac{\Delta_{13}}{2}} 
\sqrt{D - (\Delta P_{+})^2}
\right] 
\label{mixedsol1} \\
&\sin{\delta_{3,4}}& = \frac{1}{D}
\left[
-C^{(+)}
\sin {\frac{\Delta_{13}}{2}} \Delta P_{-}  
\mp 
C^{(-)}
\cos {\frac{\Delta_{13}}{2}} 
\sqrt{D - (\Delta P_{-})^2}
\right] 
\label{mixedsol2} \\
&\theta_{1,2}& =  
\frac{1}{2 D^{(-)}}
\left[
\frac{\sqrt{(P - P_{\odot}) X_{+}}}{Y_{+}} +  
\frac{\sqrt{(\bar{P} - P_{\odot}) X_{-}}}{Y_{-}}  +
\sin{\delta_{1,2}} \sin {\frac{\Delta_{13}}{2}} 
\right] 
\label{mixedsol3} \\
&\theta_{3,4}& =  
\frac{1}{2 D^{(-)}}
\left[
\frac{\sqrt{(P - P_{\odot}) X_{-}}}{Y_{-}} +  
\frac{\sqrt{(\bar{P} - P_{\odot}) X_{+}}}{Y_{+}} -
\sin{\delta_{3,4}} \sin {\frac{\Delta_{13}}{2}} 
\right] 
\label{mixedsol4} 
\end{eqnarray}
where 
\begin{eqnarray}
D &\equiv& 
{C^{(+)}}^2 
\sin^2{\left( \frac{\Delta_{13}}{2} \right)} + 
{C^{(-)}}^2 
\cos^2{\left( \frac{\Delta_{13}}{2} \right)} \\
C^{(\pm)} &\equiv& 
\frac{1}{2} \left(
\frac{Y_{+}}{X_{+}} \mp \frac{Y_{-}}{X_{-}}
\right) \\
D^{(\pm)} &\equiv& 
\frac{1}{2} \left(
\frac{X_{+}}{Y_{+}} \mp \frac{X_{-}}{Y_{-}}
\right) \\
\Delta P_{\pm} &\equiv& 
\sqrt{\frac{P - P_{\odot}}{X_{\pm}}} - 
\sqrt{\frac{\bar{P} - P_{\odot}}{X_{\mp}}} 
\end{eqnarray}
The sign in (\ref{mixedsol2}) is determined relative to 
(\ref{mixedsol1}) so that 
it reproduces the pair of degenerate solutions in the case 
of a precisely determined value of $\theta_{13}$ \cite{MNP02}.
It should be noticed that provided $\sqrt{D-(\Delta P_{\pm})^2}$
is real the constraint
$|\sin{\delta_{i}}| \leq 1$ is satisfied automatically in 
(\ref{mixedsol1}) and (\ref{mixedsol2}).

Let us focus first on the features of the same-sign 
degenerate solution. 
The set ($\theta_{i}$, $\delta_{i}$) with $i=1, 2$ ($i=3, 4$) 
describes two degenerate solutions with positive (negative) 
$\Delta m^2_{13}$ for given values of $P$ and $\bar{P}$.
Of course, they reproduce the relationships obtained by 
Burguet-Castell {\it et al} in \cite{BurguetC}:
\begin{eqnarray}
&\sin{\delta_{2}}& - \sin{\delta_{1}} = -2 
\left(
\frac{\sin{\delta_{1}} + z \cos{\delta_{1}}}{1 + z^2} 
\right) 
\label{BC1} \\
&\theta_{2} - \theta_{1}& = 
\left(
\frac{\sin{\delta_{1}} + z \cos{\delta_{1}}}{1 + z^2}
\right)
\left(
\frac{{C^{(+)}}^2 - {C^{(-)}}^2}{C^{(-)}}
\right)
\sin{\left( \frac{\Delta_{13}}{2} \right)} 
\label{BC2} 
\end{eqnarray}
where
\begin{equation}
z = \frac{C^{(+)}}{C^{(-)}}
\tan{\left( \frac{\Delta_{13}}{2} \right)}
\end{equation}

Let us illuminate how the relative phases between $\delta$'s 
between these degenerate solutions can be obtained in a 
transparent way. Toward the goal we first calculate 
$\cos{\delta_{i}}$. 
$\cos{\delta_{1,2}}$ and $\cos{\delta_{3,4}}$ 
can be obtained from (\ref{mixedsol1}) and 
(\ref{mixedsol2}), respectively, 
by replacing of $C^{(\pm)}$ by $\mp C^{(\mp)}$ 
and $\sin{\frac{\Delta_{13}}{2}}$ by 
$\cos{\frac{\Delta_{13}}{2}}$ and vice versa.
One can show by using these results that 
\begin{equation}
\cos{(\delta_{1} + \delta_{2})} = 
\cos{(\delta_{3} + \delta_{4})} = 
\frac{1 - z^2}{1 + z^2},
\end{equation}
which implies that
\begin{eqnarray}
\delta_{2} &=& \pi - \delta_{1} + \arccos((z^2-1)/(z^2+1))
\nonumber \\
\delta_{4} &=& \pi - \delta_{3} + \arccos((z^2-1)/(z^2+1)).
\end{eqnarray}
Thus in the allowed region of bi-probability space $\delta_2 ~(\delta_4)$
differs from $\pi -\delta_1 ~(\pi - \delta_3)$ by a constant,
$\arccos((z^2-1)/(z^2+1))$, which
depends on the energy and path length of the neutrino beam
but not on the mixing angle $\theta$.
Near the oscillation maximum, $z \rightarrow \infty$, this
constant vanishes so that 
$\delta_{2} \simeq \pi - \delta_{1}$ and 
$\delta_{3} \simeq \pi - \delta_{4}$
as noticed in \cite {BurguetC}.

For the mixed-sign degenerate solution one can show that 
\begin{eqnarray}
\cos{(\delta_{1} - \delta_{3})} &=& 
\frac{
\left(
\Delta P_{+} \Delta P_{-} - 
\sqrt{D - \Delta P_{+}^2} \sqrt{D - \Delta P_{-}^2}
\right)}{D}
\nonumber \\
\sin{(\delta_{1} - \delta_{3})} &=& 
\frac{
\left(
\Delta P_{+} \sqrt{D - \Delta P_{-}^2} + 
\Delta P_{-} \sqrt{D - \Delta P_{+}^2} 
\right)}{D}
\end{eqnarray}
One can show, for example, 
$\cos{(\delta_{1} - \delta_{3})} = - 1$ and 
$\sin{(\delta_{1} - \delta_{3})} = 0$
in the $\bar{P} \rightarrow P$ limit 
by noting that $\Delta P_{-} = - \Delta P_{+}$ 
in the limit. 
It means that $\delta_{3} = \delta_{1} + \pi$ (mod. $2\pi$),  
in agreement with the result obtained in Ref.~\cite{MNP02}.

The conditions for existence of the same-sign solution are
\begin{equation}
D - (\Delta P_{+})^2 \geq 0 \hskip 0.5cm \mbox{and} \hskip 0.5cm 
D - (\Delta P_{-})^2 \geq 0
\label{condition}
\end{equation}
for positive and negative $\Delta m^2_{13}$, respectively.
The condition for existence of the mixed-sign solution is 
the intersection of the two regions which satisfy the conditions 
of eq.~(\ref{condition}).
An example of the regions satisfying conditions for existence of the 
same-sign as well as mixed-sign solutions are depicted 
in Fig.~1.

The maximum value of P and  
$\bar{P}$ which allows mixed sign solutions is determined by
$D - (\Delta P_{+})^2 = D - (\Delta P_{-})^2 = 0$ with $P=\bar{P}$.
This occurs for a critical value of P given by
\begin{equation}
P_{crit} = P_\odot + { X_+ X_- D \over (\sqrt{X_+} - \sqrt{X_-})^2}
\label{Pcrit}
\end{equation}
which can be used to determine the critical value of 
$\theta$ as
\begin{equation}
\theta_{crit} = { C^{(+)} \sin^2 \frac{\Delta_{13}}{2} 
\over 2 D^{(-)} \sqrt{D} }.
\label{thetacrit}
\end{equation}
There is no degeneracy in the value of $\theta$ at this critical point,
{\it i.e.} $\theta_1 = \theta_2 = \theta_3 = \theta_4$.
An example of this can be seen in Fig.~\ref{CP-NUMI}.
At the first peak in the oscillation probability, $\Delta_{13} = \pi$,
the value of the critical $\theta$ is simply given by
\begin{equation}
\theta_{crit}(\Delta_{13}=\pi) = 
{ |Y_+| \over \sqrt{X_+} ( \sqrt{X_+} - \sqrt{X_-})}.
\end{equation}
As $\Delta_{13} \rightarrow 2 \pi$, the critical $\theta$ goes to zero
and so does the oscillation probabilities P and $\bar{P}$ as this is
the position of the first trough in the oscillation probabilities.

In Fig.~6 thru 8 we have plotted the fractional difference,
$\left(\Delta \theta \over \bar{\theta} \right)$, 
see eq.~(\ref{FracDiff}),
for the 
CERN-Frejius, JHF-SK and FNAL-NUMI possible experiments using
$\nu_\mu \rightarrow \nu_e$ and its CP-conjugate 
$\bar{\nu}_\mu \rightarrow \bar{\nu}_e$.
The regions where this fractional difference is small are regions
where the parameter degeneracy inherent in such measurements is only
important once the experimental resolution on $\theta$ for a 
fixed solution is of the same size or smaller.
Notice that near the boundaries on the allowed region
the fractional differences are small for the same sign solutions.
For the mixed sign, (1,3), fractional difference plots there is
a zero along the diagonal.
This is explained by the fact that in our 
approximation the positive and negative 
$\Delta m^2_{13}$ ellipses, for a given $\theta$,
intersect along the diagonal.
Fig.~9 thru 11 are similar to Fig.~6 thru 8 expect that the size
of solar $\Delta m^2_{12}$ has been increased by a factor of two to
$1 \times 10^{-4}$eV$^2$.
Notice that the parameter degeneracies problem is more pronounced as
the solar $\Delta m^2$ is increased.
At even larger values of  $\Delta m^2_{12}$, 
our approximations become less reliable.

\begin{figure}[t]
\vspace*{15cm}
\includegraphics{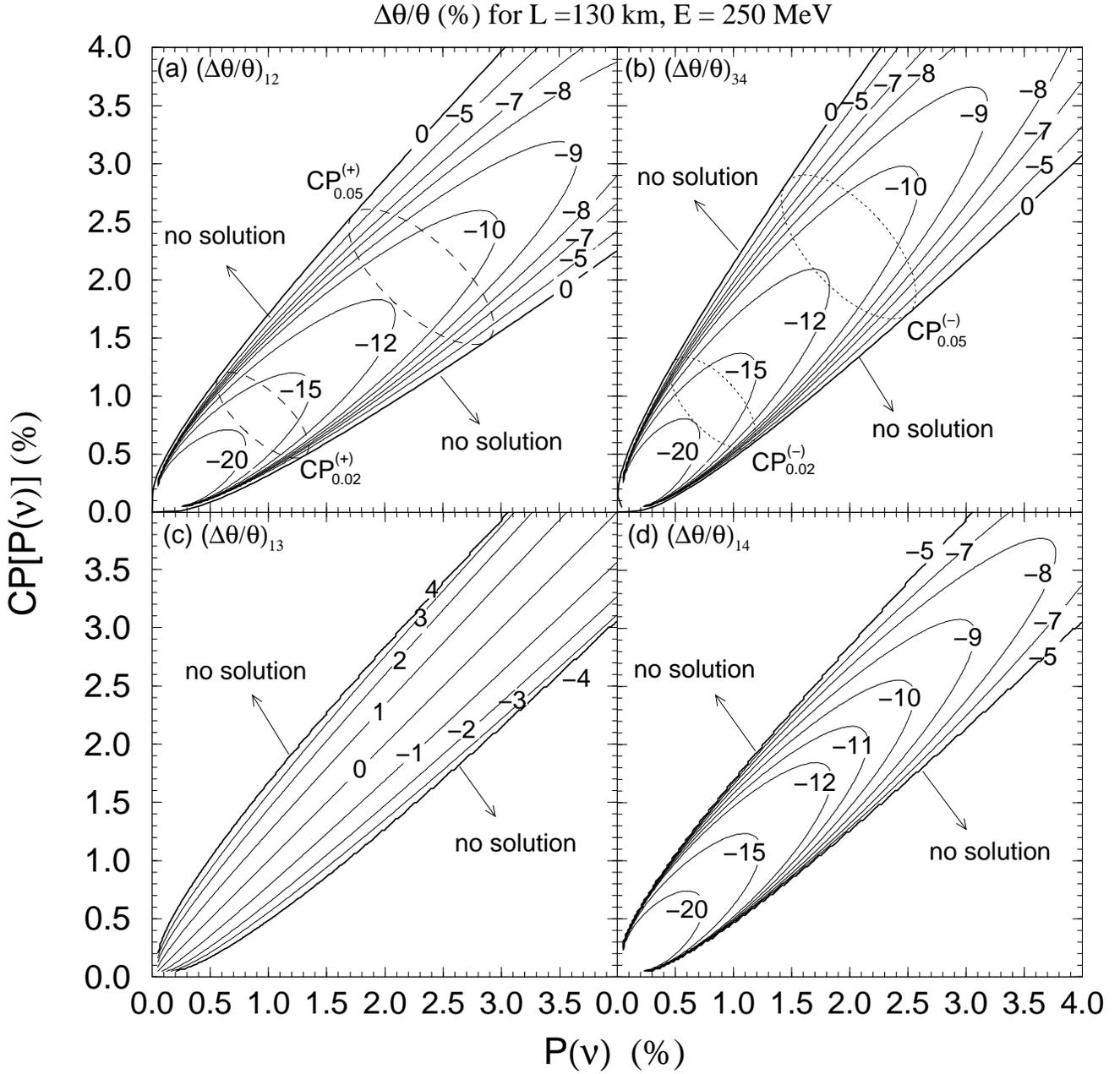}
\vspace{1.0cm}
\caption[]{
The iso-fractional differences, as a \%, for the allowed solutions
for the mixing angle $\theta_{13}$ in the 
$P(\nu) \equiv P(\nu_\mu \rightarrow \nu_e)$ verses 
$CP[P(\nu)] \equiv P(\bar{\nu}_\mu \rightarrow \bar{\nu}_e)$ plane
for the CERN-Frejius project.
The parameters are the same as in Fig.~\ref{CPegfig}
with $\Delta m^2_{12} = 5 \times 10^{-5}$ eV$^2$.
The dashed ellipses are labelled $CP^{(\pm)}_{\sin^2 2 \theta_{13}}$
to show the relevant
size of $\sin^2 2 \theta_{13}$ for this figure.
}
\label{CP-CERN}
\end{figure}

\begin{figure}[t]
\vspace*{15cm}
\includegraphics{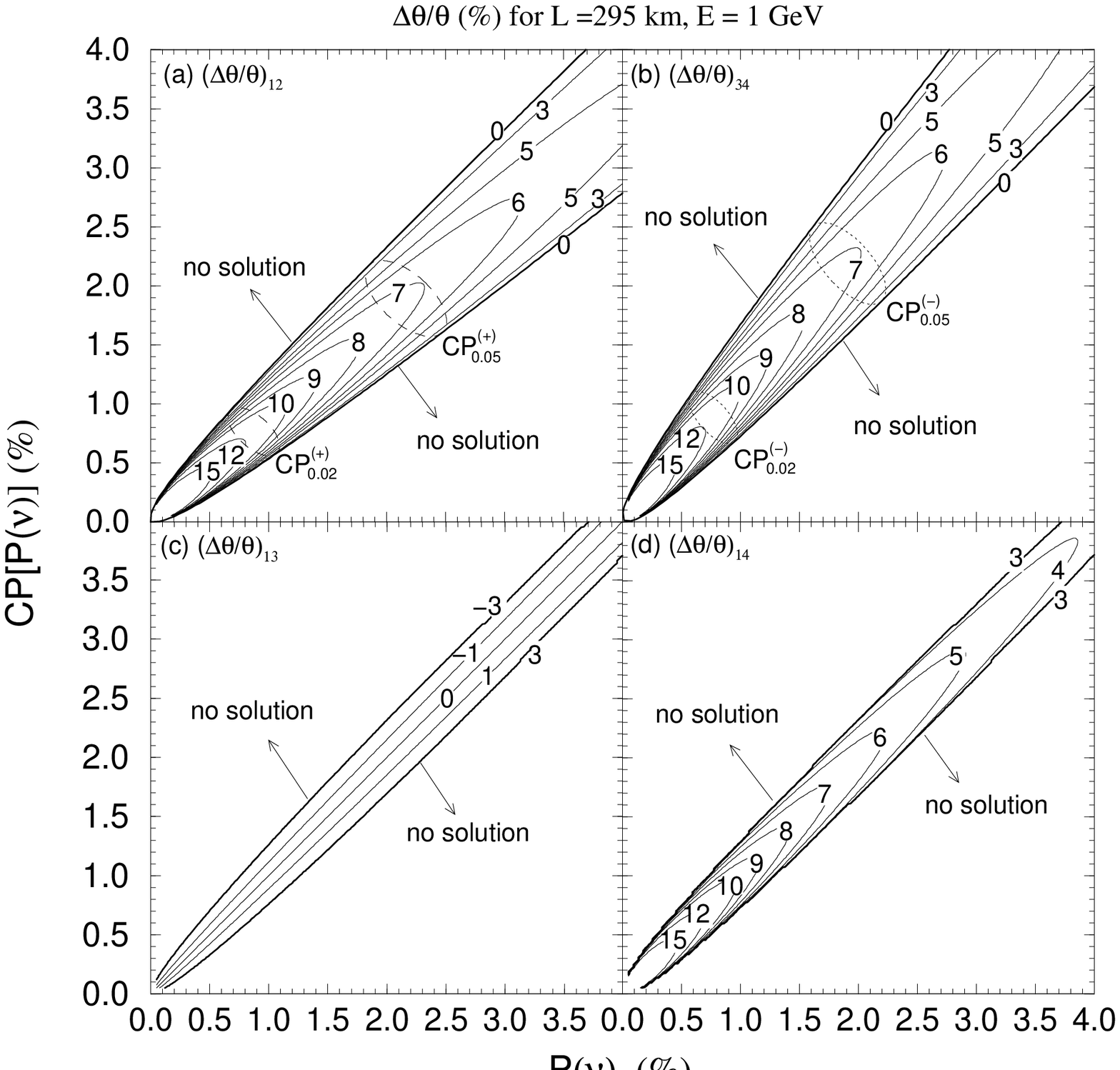}
\vspace{1.0cm}
\caption[]{
The iso-fractional differences, as a \%, for the allowed solutions
for the mixing angle $\theta_{13}$ in the 
$P(\nu) \equiv P(\nu_\mu \rightarrow \nu_e)$ verses 
$CP[P(\nu)] \equiv P(\bar{\nu}_\mu \rightarrow \bar{\nu}_e)$ plane
for the JHF-SK project.
The parameters are the same as in Fig.~\ref{CPegfig}
with $\Delta m^2_{12} = 5 \times 10^{-5}$ eV$^2$.
The ellipses are labelled as in Fig.\ref{CP-CERN}.}
\label{CP-JHF}
\end{figure}

\begin{figure}[t]
\vspace*{16cm}
\includegraphics{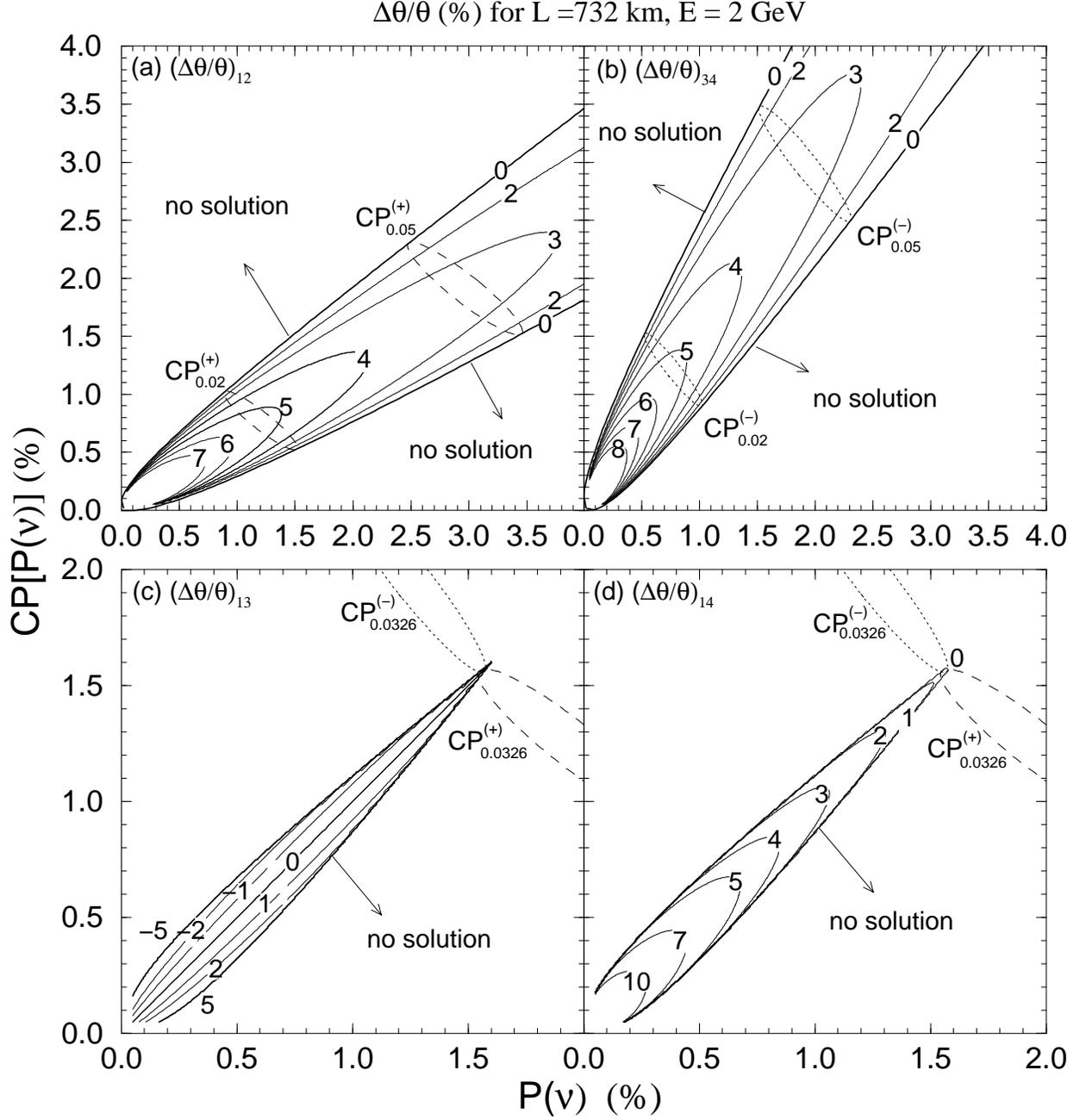}
\vspace{1.0cm}
\caption[]{
The iso-fractional differences, as a \%, for the allowed solutions
for the mixing angle $\theta_{13}$ in the 
$P(\nu) \equiv P(\nu_\mu \rightarrow \nu_e)$ verses 
$CP[P(\nu)] \equiv P(\bar{\nu}_\mu \rightarrow \bar{\nu}_e)$ plane
for the FNAL-NUMI project.
The mixed sign fractional differences,
$\left( {\Delta \theta / \bar{\theta}}\right)_{13 ~{\rm and} ~14}$ 
terminate at around 
$P=CP[P]\approx 1.6\%$ because above this probability
the sign of $\Delta m^2_{13}$ is determined, as discussed in the text.
The critical value of P ($\approx$ 1.6\%) and 
$\sin^2 2\theta_{13}$ ($\approx$ 0.033) can be calculated
from eq.~(\ref{Pcrit}) and (\ref{thetacrit}).
The parameters are the same as in Fig.~\ref{CPegfig}
with $\Delta m^2_{12} = 5 \times 10^{-5}$ eV$^2$.
The ellipses are labelled as in Fig.\ref{CP-CERN}.}
\label{CP-NUMI}
\end{figure}

\begin{figure}[t]
\vspace*{15cm}
\includegraphics{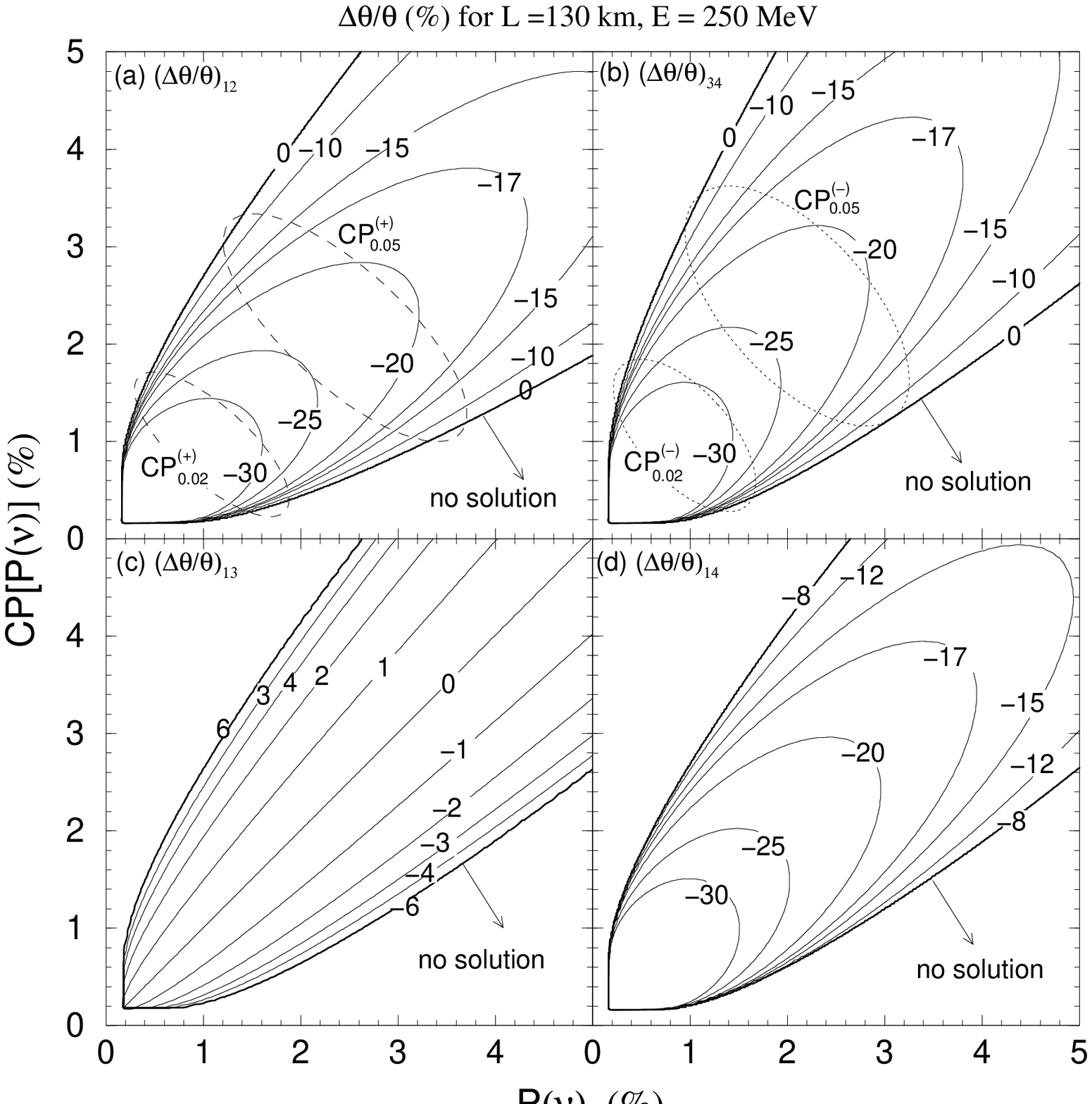}
\vspace{1.0cm}
\caption[]{
The iso-fractional differences, as a \%, for the allowed solutions
for the mixing angle $\theta_{13}$ in the 
$P(\nu) \equiv P(\nu_\mu \rightarrow \nu_e)$ verses 
$CP[P(\nu)] \equiv P(\bar{\nu}_\mu \rightarrow \bar{\nu}_e)$ plane
for the CERN-Frejius project.
The parameters are the same as in Fig.~\ref{CPegfig} except for the solar
$\Delta m^2_{12}$ which is set to $1 \times 10^{-4}eV^2$ for this plot.
The ellipses are labelled as in Fig.\ref{CP-CERN}.
}
\label{CP-CERN_hi}
\end{figure}

\begin{figure}[t]
\vspace*{15cm}
\includegraphics{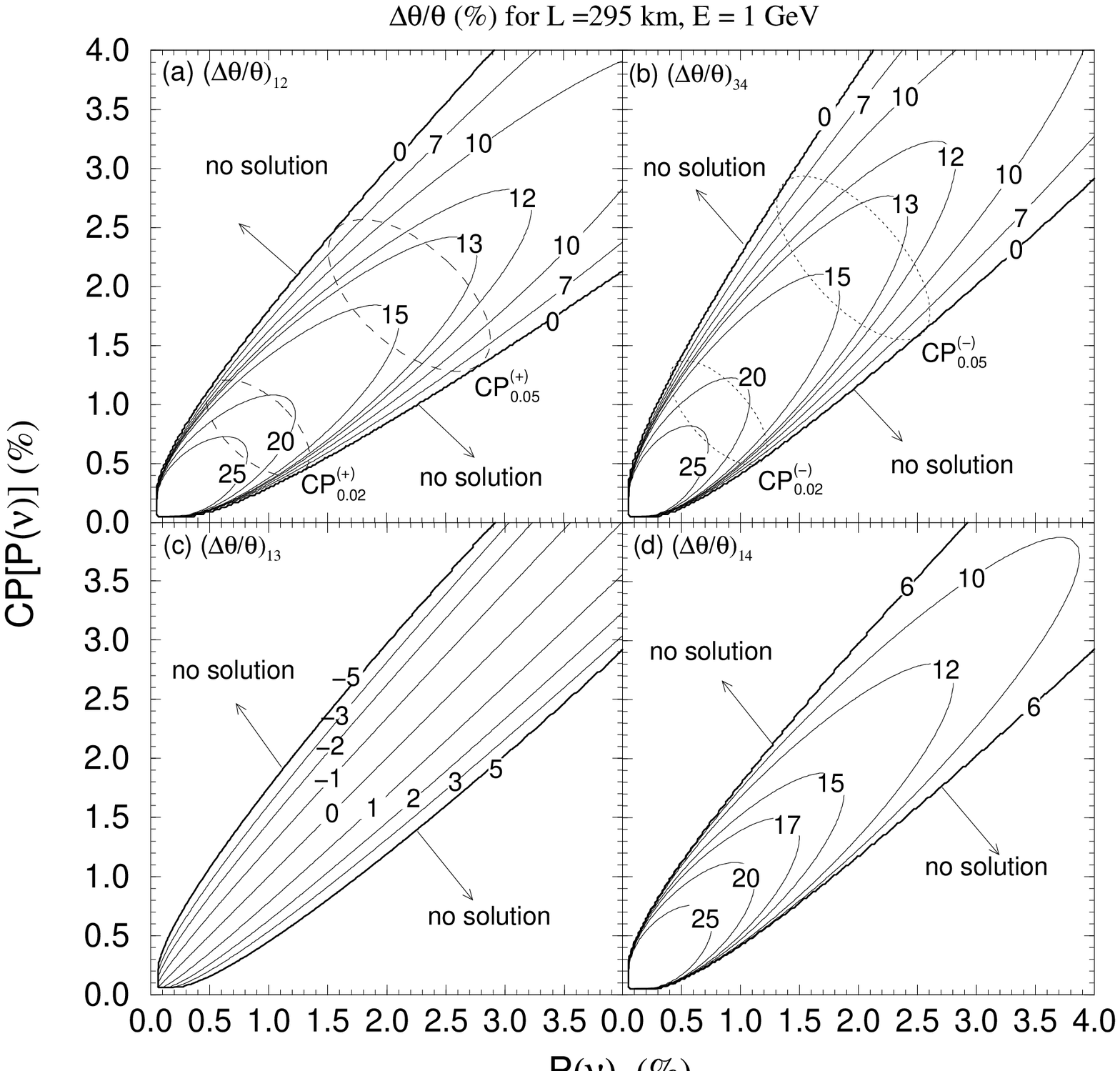}
\vspace{1.0cm}
\caption[]{
The iso-fractional differences, as a \%, for the allowed solutions
for the mixing angle $\theta_{13}$ in the 
$P(\nu) \equiv P(\nu_\mu \rightarrow \nu_e)$ verses 
$CP[P(\nu)] \equiv P(\bar{\nu}_\mu \rightarrow \bar{\nu}_e)$ plane
for the JHF-SK project.
The parameters are the same as in Fig.~\ref{CPegfig} except for the solar
$\Delta m^2_{12}$ which is set to $1 \times 10^{-4}eV^2$ for this plot.
The ellipses are labelled as in Fig.\ref{CP-CERN}.
}
\label{CP-JHF_hi}
\end{figure}

\begin{figure}[t]
\vspace*{15cm}
\includegraphics{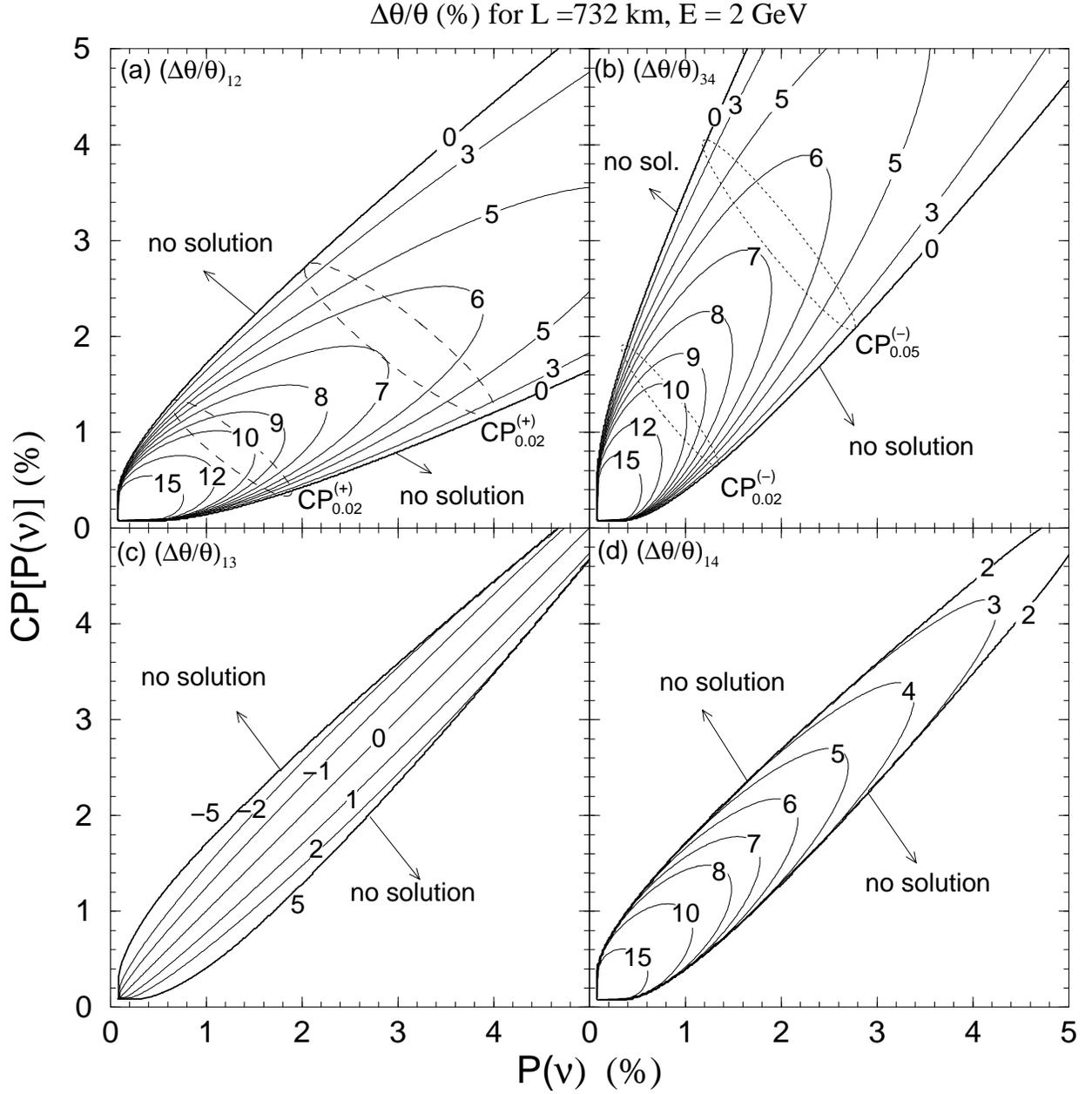}
\vspace{1.0cm}
\caption[]{
The iso-fractional differences, as a \%, for the allowed solutions
for the mixing angle $\theta_{13}$ in the 
$P(\nu) \equiv P(\nu_\mu \rightarrow \nu_e)$ verses 
$CP[P(\nu)] \equiv P(\bar{\nu}_\mu \rightarrow \bar{\nu}_e)$ plane
for the FNAL-NUMI project.
The parameters are the same as in Fig.~\ref{CPegfig} except for the solar
$\Delta m^2_{12}$ which is set to $1 \times 10^{-4}eV^2$ for this plot.
The ellipses are labelled as in Fig.\ref{CP-CERN}.
}
\label{CP-NUMI_hi}
\end{figure}

\section{Summary and Conclusion}

In this paper we have given a complete analytic treatment of the
parameter degeneracy issue for $\theta_{13}$, sign of $\Delta m^2_{13}$
and the CP and T violating phase $\delta$ that appears in 
neutrino oscillations.
For a given neutrino flavor transition probability and its CP or T
conjugate probability we have derived the allowed values of 
$\theta_{13}$, sign of $\Delta m^2_{13}$
and the CP and T violating phase $\delta$.
We have given explicit expressions of degenerate solutions and
obtained, among other things, exact formulas for the relationship between 
solutions of $\delta$'s up to the correction of order 
$\left( \frac{\Delta m^2_{12}}{\Delta m^2_{13}} \right)^2$.

In general there is a four-fold degeneracy, two allowed values
of $\theta_{13}$ for both signs of $\Delta m^2_{13}$. 
This is always true for the T violation measurement whereas for
the CP violation measurement the four-fold degeneracy 
can be reduced to two-fold degeneracy if matter effects are
sufficiently large, or we live close to the region 
$\delta \sim \pi/2$ or $3\pi/2$.  
The significance of matter effects
dependence on the energy of the neutrino beam, the
separation between the source and the detector as well as the density
of matter between them.
The fractional difference of $\theta_{13}$ between the various solutions has
been calculated which can be compared with the experimental
sensitivity for a given setup to determine whether or not
the degeneracy issue is significant or not.

For the possible future experimental setups
CERN-Frejius, JHK-SK and FNAL-NUMI
we have given numerical results for the channel
$\nu_\mu \rightarrow \nu_e$
and its CP and T conjugate.
The CP conjugate being most relevant for these future
Super-beam experiments.
For the CERN-Frejius, JHF-SK and FNAL-NUMI experimental setups
the parameter degeneracy issue is only relevant once the experimental
resolution on the determination of $\theta_{13}$ is better 
than 15\%, 10\% and 5\% respectively, assuming a transition probability
near 1\% and a $\Delta m^2_{12} = 5 \times 10^{-5} eV^2$,
see Fig. \ref{CP-CERN} - \ref{CP-NUMI}.
At larger values of $\Delta m^2_{12}$ the parameter degeneracy issue
becomes more important.
These iso-fractional difference plots are useful for comparing
the sensitivity of 
different experimental setups,
neutrino energy, path length and experimental sensitivity,
to this parameter degeneracy issue.

\acknowledgments

HM thanks Theoretical Physics Department of Fermilab for warm
hospitality extended to him through the summer visitors program in 
2002.
This work was supported by the Grant-in-Aid for Scientific Research
in Priority Areas No. 12047222, Japan Ministry
of Education, Culture, Sports, Science, and Technology.
Fermilab is operated by URA under DOE contract No.~DE-AC02-76CH03000.

\section{appendix}

The standard flavor transition probability for neutrino oscillations
in the $\nu_\mu \rightarrow \nu_e$ channel can be written as
\begin{eqnarray}
P(\nu)_{\pm} &=& X_{\pm} \theta^2 + 
Y_{\pm} \theta \cos {\left( \delta \pm \frac{\Delta_{13}}{2} \right)} + 
P_{\odot}
\end{eqnarray}
where the $\pm$ signs in $X_{\pm}$ and $Y_{\pm}$ refer to
positive or negative values of $\Delta m^2_{13}$, $\theta$ is an abbreviation
for $\sin \theta_{13}$ 
and $P_\odot$ indicates the terms related to solar neutrino oscillations.
For details on the approximations used in deriving this transition
probability see ref. \cite{golden}.
All other channels used in this paper,  
$\nu_e \rightarrow \nu_\mu$ and
$\bar{\nu}_\mu \rightarrow \bar{\nu}_e$,
can also be expressed with the same variables, see Sec. III and IV.

The coefficients $X_{\pm}$ and $Y_{\pm}$ are determined by
\begin{eqnarray}
X_{\pm} &=& 4 s^2_{23} 
\left(
\frac{\Delta_{13}}{B_{\mp}} 
\right)^2
\sin^2{\left(\frac{B_{\mp}}{2}\right)}, 
\label{X} \\
Y_{\pm} &=& \pm 8 c_{12}s_{12}c_{23}s_{23}
\left(
\frac{\Delta_{12}}{aL}
\right)
\left(
\frac{\Delta_{13}}{B_{\mp}}
\right)
\sin{\left(\frac{aL}{2}\right)}
\sin{\left(\frac{B_{\mp}}{2}\right)}
\label{Y}\\
P_{\odot} & = & c^2_{23} \sin^2{2\theta_{12}} 
\left(\frac{\Delta_{12}}{aL}\right)^2
\sin^2{\left(\frac{aL}{2}\right)}
\end{eqnarray}
with 
\begin{eqnarray}
\Delta_{ij}  \equiv \frac{|\Delta m^2_{ij}| L}{2E}
\quad
{\rm and} \quad B_{\pm} \equiv |aL \pm \Delta_{13}|,
\end{eqnarray}
where $a = \sqrt{2} G_F N_e$ denotes the index of refraction 
in matter with $G_F$ being the Fermi constant and $N_e$ a constant 
electron number density in the earth. 

Obviously from the above definitions, 
$X_\pm$ and $Y_\pm$ satisfy the identity
\begin{equation}
{ Y_+ \over \sqrt{X_+}}
= - ~{ Y_- \over \sqrt{X_-}}
\end{equation}
which is used throughout this paper.

\end{document}